\documentclass[a4paper,11pt]{article}
\pdfoutput=1 
\usepackage[english]{babel}	
\usepackage{xspace}
\usepackage{microtype}	
\usepackage[utf8]{inputenc}	

\usepackage{jheppub}  
\usepackage[T1]{fontenc}  
\usepackage{amsmath}
\usepackage{slashed}
\usepackage{graphicx}
\usepackage{pgfplots}
\usepackage{color}
\usepackage{rotating}
\usepackage{gensymb}
\usepackage{simplewick}
\usepackage{enumerate}
\usepackage{amsmath,amssymb,amsfonts,dsfont,bm}
\usepackage{subfigure}

\newcommand{\abs}[1]{\left| #1 \right|}
\newcommand{\braces}[1]{\left\lbrace #1 \right\rbrace}

\newcommand{\dd}{\mathop{}\!\mathrm{d}}

\newcommand{\andeq}{\quad \mathrm{and} \quad}
\newcommand{\Tr}{\mathop{}\!\mathrm{Tr}}

\newcommand{\U}{\mathrm{U}}
\newcommand{\SU}{\mathrm{SU}}

\newcommand{\SO}{\mathrm{SO}}

\newcommand{\udindices}[2]{\phantom{}^{#1}\phantom{}_{#2}}
\newcommand{\duindices}[2]{\phantom{}_{#1}\phantom{}^{#2}} 
\newcommand{\pole}[1]{\left. #1 \right|_{1/\epsilon}}

\usepackage{xcolor,colortbl}
\definecolor{Blue}{RGB}{140,165,195}
\definecolor{Purple}{RGB}{255,145,145}
\definecolor{orange}{cmyk}{0,0.5,1,0}
\definecolor{rossoCP3}{cmyk}{0,.88,.77,.40}
\definecolor{graa}{rgb}{0.8,0.8,0.8}
\definecolor{blaa}{rgb}{0.2,0.2,0.6}
\hypersetup{
	colorlinks, 
	bookmarksopen, 
	bookmarksnumbered,
	citecolor=blaa, 		
	linkcolor=rossoCP3,	
	urlcolor=rossoCP3,			
}

\begin{document}

\title{\boldmath \textcolor{rossoCP3} {Gauge-Yukawa theories: Beta functions at large $ N_f$}}

\author[a]{Oleg Antipin,}
\author[b]{Nicola Andrea Dondi,}
\author[b]{Francesco Sannino,}
\author[b]{Anders Eller Thomsen,}
\author[b,c]{Zhi-Wei Wang,}

\affiliation[a]{Rudjer Boskovic Institute, Division of Theoretical Physics,\\ Bijeni\v cka 54, HR-10000 Zagreb, Croatia}
\affiliation[b]{$\rm{CP}^3$-Origins, University of Southern Denmark,\\Campusvej 55
5230 Odense M, Denmark}
\affiliation[c]{Department of Physics, University of Waterloo,\\Waterloo, ON, N2L 3G1, Canada}

\emailAdd{oantipin@irb.hr}
\emailAdd{dondi@cp3.sdu.dk}
\emailAdd{sannino@cp3.sdu.dk}
\emailAdd{aethomsen@cp3.sdu.dk}
\emailAdd{wang@cp3.sdu.dk}

\abstract{We consider the dynamics of gauge-Yukawa theories in the presence of a large number of   matter constituents.   
We first review the current status for the renormalization group equations of gauge-fermion theories and extend the results to semi-simple groups. In this regime these theories develop an interacting ultraviolet fixed point that for the semi-simple case leads to a rich phase diagram. The latter contains a complete asymptotically safe fixed point repulsive in all couplings.  We then add two gauged Weyl fermions belonging to arbitrary representations of the semi-simple gauge group and a complex, gauged scalar to the original gauge-fermion theory allowing for new Yukawa interactions and quartic scalar self-coupling. Consequently, we determine  the first nontrivial order in $ 1/N_f $ for the Yukawa and quartic beta functions. Our work elucidates, consolidates and extends results obtained earlier in the literature.  We also acquire relevant knowledge about the dynamics of gauge-Yukawa theories beyond perturbation theory.  Our findings are applicable to any extension of the standard model featuring a large number of fermions such as asymptotic safety.  
\\[1em]
	{\it \small Preprint: CP$^3$-Origins-2018-011}
 }

\maketitle
\flushbottom

\section{Introduction}

The most general class of four-dimensional, renormalizable quantum field theories (QFTs) are in the form of gauge-Yukawa theories. Their dynamics underlies the Standard Model (SM) interactions  and those of  any of its sensible  extensions. It is therefore paramount  to gain a deeper understanding of their dynamics which is often limited to perturbation theory. 

Fundamental theories are those gauge-Yukawa theories that, according to Wilson~\cite{Wilson:1971bg,Wilson:1971dh}, are well defined at arbitrarily short distances. Asymptotically free~ \cite{Gross:1973ju,Politzer:1973fx}  and safe \cite{Litim:2014uca} QFTs are complementary examples of fundamental theories\footnote{The instanton analysis and contribution can be found in \cite{Sannino:2018suq}.}. The recent discovery of four-dimensional, controllable, in the perturbative sense, asymptotically safe QFTs~\cite{Litim:2014uca, Litim:2015iea} has opened the way to novel dark and bright extensions of the SM~\cite{Sannino:2015sel,Sannino:2014lxa,Abel:2017ujy,Abel:2017rwl,Pelaggi:2017wzr,Mann:2017wzh,Pelaggi:2017abg,Bond:2017wut}. More generally, it is interesting to investigate the short distance fate of the SM and its extensions including gravity~\cite{Eichhorn:2018vah,Eichhorn:2017muy,Reichert:2017puo,Eichhorn:2017sok,Eichhorn:2017lry}. 

To gain information beyond perturbation theory, one can use supersymmetry. A systematic investigation of non-perturbative constraints that  a supersymmetric, asymptotically safe QFT must abide, including a-maximization \cite{Intriligator:2003jj} and collider bounds \cite{Hofman:2008ar}, appeared in \cite{Intriligator:2015xxa} extending and correcting the results of \cite{Martin:2000cr}. Building upon results of \cite{Intriligator:2003jj}, the first evidence for non-perturbative, supersymmetric safety was  gathered in \cite{Bajc:2016efj}  and further analyzed in \cite{Bajc:2017xwx}. 

Non-perturbative results can also be deduced for non-supersymmetric theories when considering specific  limits in theory space. For example, building upon the large $N_f$ results of \cite{PalanquesMestre:1983zy,Gracey:1996he,Holdom:2010qs,Pica:2010xq,Shrock:2013cca}, that gauge-fermion theories at any finite number of colors can be argued to develop a non-perturbative ultraviolet (UV) fixed point \cite{Antipin:2017ebo}. Consequently, one can extend the original conformal window, reviewed in \cite{Sannino:2009za,Pica:2017gcb}, to include an asymptotically safe phase \cite{Antipin:2017ebo}.

It is therefore timely to consider the dynamics of gauge-Yukawa theories at large $N _f$ \cite{Kowalska:2017pkt,Pelaggi:2017abg,Ferreira:1997rc,Ferreira:1997bi}. We investigate this here by elucidating, consolidating, and extending the results obtained earlier in the literature. The results are useful when searching for asymptotically safe extensions of the SM~\cite{Mann:2017wzh,Pelaggi:2017abg}.

The paper is organized as follows. In section \ref{sec:model}, we will introduce our model and the renormalization conventions used throughout the paper. Section \ref{sec:gauge_fermion} then proceeds to review the current status of large $ N_f $ computations for the gauge beta function. We then generalize the results to semi-simple gauge groups and in addition we present their phase diagrams. In section \ref{sec:SelfishYukawa}, we provide a detailed computation of the Yukawa and quartic coupling beta functions at the first nontrivial order in $ 1/N_f $ for generic, semi-simple gauge groups. Section \ref{sec:conclusion} concludes the paper. The explicit derivations of the various resummation formulas used in the large $ N_f $ computations can be found in the appendix \ref{sec:appendix}.

\section{Gauge-Yukawa models: notation and conventions} \label{sec:model} 
We consider both Abelian and non-Abelian semi-simple gauge-Yukawa models featuring $ N_f $ vector-like fermions, $ \Psi_I $  charged under the full gauge group.  Additionally,  the models contain two Weyl spinors, $ \chi$ and $ \xi $, and a complex scalar, $ \phi $, such that there is enough content to form a (chiral) Yukawa coupling  among these three fields, and for quartic scalar self-interactions to emerge\footnote{Gauge anomalies are  avoided by either adding new chiral fermions or by arranging  $ \chi, \xi $  in anomaly free representations of the gauge group. Our results are adaptable to a given gauge-anomaly free model. }. The field content of the model is summarized in Table~\ref{tab:field_content} where we report the transformation of each matter field with respect to the gauge interactions. 
 
\begin{table}[t]
\[ \begin{array}{|c|cc|cc|} \hline
{\rm Fields} & \SO(1,3)^{+} & \SU(N_f)  & \U(1) & \times_\alpha G_\alpha  \\ \hline 
\Psi & \left(\frac{1}{2},0\right) \oplus \left(0,\frac{1}{2}\right)  & N_f  & q_\Psi & \otimes_\alpha R_\Psi^\alpha   \\
\chi & \left(\frac{1}{2},0\right) &1 & q_\chi & \otimes_\alpha R_\chi^\alpha   \\
\xi & \left(\frac{1}{2},0\right) & 1 & q_\xi & \otimes_\alpha R_\xi^\alpha   \\ 
\phi  & \left(0,0\right) & 1 & q_{\phi} & \otimes_\alpha R_\phi^\alpha  \\    
  \hline
     \end{array} 
\]
\caption{Summary of the field content of the model. The first two columns detail the transformation of each field under Lorentz and flavor symmetry. $ q_{\Psi, \chi, \xi, \phi} $ denotes the $ \U(1) $ charges of the fields, while $R_{\Psi, \chi, \xi, \phi}^\alpha $ are the representation of the fields under each simple gauge group labeled by $\alpha$.}
\label{tab:field_content}
\end{table} 

The Lagrangian of the theory reads 
	\begin{equation}\begin{split} \label{eq:L_model}
	\mathcal{L} = -\tfrac{1}{4} F^{A}_{\mu\nu} F^{A,\mu\nu}& + \sum_{I=1}^{N_f} i \overline{\Psi}_{I} \gamma^{\mu} D_{\mu} \Psi^{I} + i \bar{\chi} \bar{\sigma}^{\mu} D_{\mu} \chi + i \bar{\xi} \bar{\sigma}^{\mu} D_{\mu} \xi + (D_\mu \phi )^\dagger (D_\mu \phi )\\
	& -  \left(  y_{aij} \phi^{a} \chi^{i} \xi^{j} + y^{\ast, aij} \phi^{\ast}_a \bar{\chi}_i \bar{\xi}_j \right)  - \tfrac{1}{4} \lambda\udindices{ab}{cd} \phi^{\ast}_a \phi^{\ast}_b \phi^c \phi^d \ ,
	\end{split}\end{equation}
where the index $I=1,\cdots N_f$ is the $\Psi$ flavor index, $i,j$ are gauge indices for $\chi$ and $\xi$,  and $a, b, c, d $ are reserved for the gauged  scalar indices within a given representation that can be read off from the associated covariant derivative 
\begin{equation}\begin{split}
D_{\mu}\Psi_I^i = \left[\partial_{\mu} + i g   A^A_{\mu} \,(T_{\Psi}^A)\udindices{i}{j} \right] \Psi_I^j \ ,\\
D_{\mu}\chi^i = \left[\partial_{\mu} + i g   A^A_{\mu} \,(T_{\chi}^A)\udindices{i}{j} \right] \chi^j  \ ,\\
D_{\mu}\xi^i = \left[\partial_{\mu} + i g  A^A_{\mu} \, (T_{\xi}^{A})\udindices{i}{j} \right] \xi^j  \ ,\\
D_{\mu}\phi^a = \left[\partial_{\mu} + i g A^A_{\mu} \,(T_{\phi}^{A})\udindices{a}{b}  \right] \phi^b  \ .\\
\end{split}\end{equation}
In the most general version of the model, the gauge group is allowed to be semi-simple. The generalization of the covariant derivative in this case is straight forward. Gauge invariance imposes the following constraints
\begin{equation}\begin{split}
0 &= y_{bij} (T_{\phi}^A)\udindices{b}{a} + y_{akj} (T_{\chi}^{A})\udindices{k}{i} + y_{aik} (T_{\xi}^{A})\udindices{k}{j} \ , \\
0 &= -\lambda\udindices{eb}{cd} (T_{\phi}^{\ast A})\duindices{e}{a} -\lambda\udindices{ae}{cd} (T_{\phi}^{\ast A})\duindices{e}{b} + \lambda\udindices{ab}{ed} (T_{\phi}^A)\udindices{e}{c}+ \lambda\udindices{ab}{ce} (T_{\phi}^A)\udindices{e}{d} \ , \\
\end{split}\end{equation}
while the constraint on the Abelian charges reads 
	\begin{equation}
	q_{\phi} + q_\chi + q_\xi =0.
	\end{equation} 

\bigskip
To prepare for the large number of flavors limit, the gauge couplings for each gauge group $G_\alpha$ are rescaled as follows
	\begin{equation}\label{eq:K_coupling}
	K_{\alpha} = \dfrac{g_\alpha^2 \mathcal{N}\,S_2(R_\Psi^{\alpha})}{4\pi^2\, d(R_\Psi^{\alpha})}, \qquad \mathrm{where} \quad \mathcal{N} = N_f \prod_{\alpha} d(R_\Psi^{\alpha}).
	\end{equation}
Here the Dynkin index $S_2(R_\Psi^{\alpha})$ is defined via the relation $S_2(R_\Psi^{\alpha}) \delta^{AB} = \Tr\left[ T_{R_\Psi^{\alpha}}^{A} T_{R_\Psi^{\alpha}}^{B} \right] $. In the fundamental representation of an $\SU(N)$ group we take it to assume the value $1/2$.  The dimension of a given representation is indicated with $d(R_\Psi^{\alpha})$. 

\subsection{Renormalization conventions}
We will now briefly summarize our renormalization conventions to prepare for the computations of the RG-functions in the model. We denote all bare fields and couplings with subscript $ 0 $.

In the Lagrangian   \eqref{eq:L_model}, the bare fields  renormalize according to:
	\begin{gather}
	A_{\alpha,0}^\mu = Z_{A_\alpha}^{1/2}\mu^{-\epsilon/2} A_\alpha^{\mu}\ , \quad \Psi_0 = Z_\Psi^{1/2} \mu^{-\epsilon/2} \Psi\ , \quad  \phi_0 = Z_\phi^{1/2} \mu^{-\epsilon/2} \phi \ , \nonumber \\
	\chi_{0} = Z_\chi^{1/2} \mu^{-\epsilon/2} \chi\ , \quad \xi_{0} = Z_\xi^{1/2} \mu^{-\epsilon/2} \xi \ ,
	\end{gather}
while the bare couplings are given by 
	\begin{equation} \label{eq:bare_y_and_lambda}
	\begin{split}
	y_{0,aij} &= \left(Z_\chi Z_\xi Z_\phi \right)^{-1/2} \mu^{\epsilon/2} (y_{aij} + \delta y_{aij}) \ , \\
	\lambda_0\udindices{ab}{cd} &= Z_\phi^{-2} \mu^{\epsilon} (\lambda\udindices{ab}{cd} + \delta \lambda\udindices{ab}{cd}) \ , \\ 
	g_{\alpha,0} & = \tilde{g}_{\alpha,0} \mu^{\epsilon/2} = Z_{K_\alpha}^{-1/2} \mu^{\epsilon/2} g_\alpha \ .
	\end{split}
	\end{equation} 
We use dimensional regularization with $d=4-\epsilon$. The field renormalizations are expanded in terms of their $ \epsilon $ poles, writing 
	\begin{equation}
	Z_i = 1 + \sum_{k=1}^{\infty} \dfrac{1}{\epsilon^{k} } Z_i^{(k)}.
	\end{equation} 
Similarly, the counter terms are expressed as
	\begin{equation} \label{eq:generic_counter_terms}
	\delta y_{aij} = \sum_{k=1}^{\infty} \dfrac{1}{\epsilon^{k}} \delta y^{(k)}_{aij} \andeq  \delta \lambda\udindices{ab}{cd} = \sum_{k=1}^{\infty} \dfrac{1}{\epsilon^{k}} \delta \lambda^{(k)} \udindices{ab}{cd}.
	\end{equation}
It is now possible to expresses the beta function for the couplings in terms of the field-strength renormalizations and the counter terms of the renormalized Lagrangian in the above notation. The beta functions, $ \beta_x = \dd x /\dd \ln \mu $, are given by
	\begin{equation}\label{eq:beta_formulas}
	\begin{split}
	\beta_{y,aij} &= \left(-\dfrac{1}{2} + K_\beta \dfrac{\partial}{\partial K_\beta} + \dfrac{y_{ekl}}{2} \dfrac{\partial}{\partial y_{ekl}} + \lambda\udindices{ef}{gh} \dfrac{\partial}{\partial \lambda\udindices{ef}{gh}} \right) \left[ \delta y^{(1)}_{aij} - \dfrac{Z_\chi^{(1)} + Z_\xi^{(1)} + Z_\phi^{(1)}}{2} y_{aij} \right], \\
	\beta_\lambda\udindices{ab}{cd} &= \left(-1 + K_\beta \dfrac{\partial}{\partial K_\beta} + \dfrac{y_{ekl}}{2} \dfrac{\partial}{\partial y_{ekl}}+  \lambda\udindices{ef}{gh} \dfrac{\partial}{\partial \lambda\udindices{ef}{gh}} \right) \left[\delta\lambda^{(1)} \udindices{ab}{cd} - 2 Z_\phi^{(1)} \lambda\udindices{ab}{cd} \right], \\
	\beta_{K_\alpha} &= \left(-1 + K_\beta \dfrac{\partial}{\partial K_\beta} + \dfrac{y_{ekl}}{2} \dfrac{\partial}{\partial y_{ekl}} + \lambda\udindices{ef}{gh} \dfrac{\partial}{\partial \lambda\udindices{ef}{gh}} \right) \left[- Z_{K_\alpha}^{(1)} K_\alpha \right].
	\end{split}
	\end{equation}

In order to practically evaluate the gauge field renormalization we make use of
	\begin{equation}\label{eq:gauge_field_renormalization}
	0 = \mathrm{div}_\epsilon \left[Z_A (1-\Pi_B(\{ x_0 \})) \right] ,
	\end{equation}
where $\{ x_0 \}$ represents the full set of bare couplings and $\Pi_B$ is the bare, 1PI, 2-point function of the gauge bosons after having factorized out momenta and polarization structure; $i\Pi_{B,\mu\nu}(p)=i p^2 \Delta_{\mu\nu}(p) \Pi_B(p^2)$ with  $ \Delta_{\mu\nu}(p) = \eta_{\mu\nu} - p_\mu p_\nu/p^2 $. 
Similarly, to compute the fermion and scalar field renormalization, we rely on the following relations involving the bare, 1PI, 2-point fermion, $-i \Sigma_B(p) $ and 2-point scalar, $ -i S_B(p^2) $, functions:
	\begin{equation}\begin{split}
	0 &= \mathrm{div}_\epsilon \left[Z_{\chi,\xi} \left(1- \dfrac{\dd}{\dd \bar{\sigma} \cdot p} \Sigma_{\chi(\xi),B}( \{ x_0 \} \right) \right] ,\\
	0 &= \mathrm{div}_\epsilon \left[Z_\phi \left(1- \dfrac{\dd}{\dd p^2} S_B(\{ x_0 \} )\right) \right] .
	\end{split}\end{equation}
Finally for the renormalization of the couplings we employ 
	\begin{equation}\begin{split}
	0 &= \mathrm{div}_\epsilon \left[ Z_\phi^{1/2} Z_\chi^{1/2} Z_\xi^{1/2} Y_B ( \{ x_0 \} ) \right] \\
	0 &= \mathrm{div}_\epsilon \left[ Z_\phi^{2}  \Lambda_B ( \{ x_0 \} ) \right]
	\end{split}\end{equation} 
where $iY_B\, , i\Lambda_B$ are the bare, 1PI, 3- and 4-point functions.  These are used to renormalize the Yukawa and quartic couplings.

\section{Gauge-fermion theory} \label{sec:gauge_fermion}
We start with reviewing the large ${\cal N}$  dynamics in gauge-fermion theory investigated some time ago in \cite{PalanquesMestre:1983zy,Gracey:1996he,Holdom:2010qs,Pica:2010xq,Shrock:2013cca}. This means that we drop $\phi$, $\chi$, and $\xi$ from the beginning. We will extend the analysis to include semi-simple gauge groups. The full dynamics including $\phi$, $\chi$, and $\xi$ will be investigated in Section \ref{sec:SelfishYukawa}. 

Only a limited set of diagrams contribute when computing the RG functions in the large $\cal{N}$ limit. In general the order, $(1/\mathcal{N})^{k}$, of a diagram in the large $ \mathcal{N} $ expansion can be determined as
	\begin{equation}
	k = \text{ powers of $g_0^2$}  \: - \: \text{\# of fermion loops}.
	\end{equation}
It follows that dressing gauge lines with  $ \Psi $ fermion bubbles (a bubble chain) does not increase the order of a diagram. To obtain the contribution at a given order in $1/\mathcal{N}$, it is sufficient to consider a small set of diagrams, but one has to sum over every number of bubbles on each gauge line. The resulting power series in $K$ is so well behaved that it is often possible to obtain a closed form expression for the $1/\epsilon$ pole. 

In the following computations we will need to have an expression for the bubble chain. Each elementary bubble stems from a bare, 1PI, $ \Psi $-fermion loop that in $\overline{\text{MS}}$  reads   
	\begin{equation} \label{eq:one_bubble_propagator}
	\begin{split}
	&i \Pi_{\mu\nu}(p) = i p^2 \Delta_{\mu\nu}(p) \Pi_0(p^2), \qquad \mathrm{where}\\
	&\Pi_0(p^2) = - 2 K_0 \Gamma_0(\epsilon) \left(-\dfrac{4\pi \mu^2}{p^2}\right)^{\epsilon/2} \andeq \Gamma_0(\epsilon) =  \dfrac{\Gamma^2(2-\tfrac{\epsilon}{2}) \Gamma(\tfrac{\epsilon}{2})}{\Gamma(4-\epsilon)} .
	\end{split}
	\end{equation}
Note that  $K_0$ is related to $\tilde{g}_0$ like $K_{\alpha}$ is related to $g_{\alpha}$ in  \eqref{eq:K_coupling} . To avoid making the notation heavy we dropped a tilde on $K_0$.

The expression for a bubble chain with $ n > 0 $ bubbles and $ n + 1 $ free gauge propagators, $ D_{\mu \nu}(p)$, reduces to 
	\begin{equation} \label{eq:n_bubble_propagator}
	\begin{split}
	D^{(n)}_{\mu\nu}(p) &= \left[ D_{\mu\mu_1}(p) i\Pi^{\mu_1 \mu_2}(p) \right] \left[ D_{\mu_2\mu_3}(p) i\Pi^{\mu_3 \mu_4}(p) \right] \cdots D_{\mu_{2n}\nu}(p) \\
	&= \dfrac{-i}{p^2} \Delta_{\mu\nu}(p) \, \Pi_0^n(p^2) \ .
	\end{split} 
	\end{equation}	
The chain is fully transverse in $p$  because the gauge-fixing parameter does not renormalize in $\overline{\text{MS}}$. In our computations we  work in the Landau (Lorenz) gauge. This has the added benefit that $ D^{(n=0)}_{\mu \nu}(p) = D_{\mu \nu}(p)$. The discussion above applies to each individual gauge group $\alpha$.

 \subsection{Large $\mathcal{N}$ gauge beta function}
\begin{figure}[t]
\centering
\subfigure[]{
\label{bubble_1}
\begin{minipage}{6cm}
\centering
\includegraphics[width=1\columnwidth]{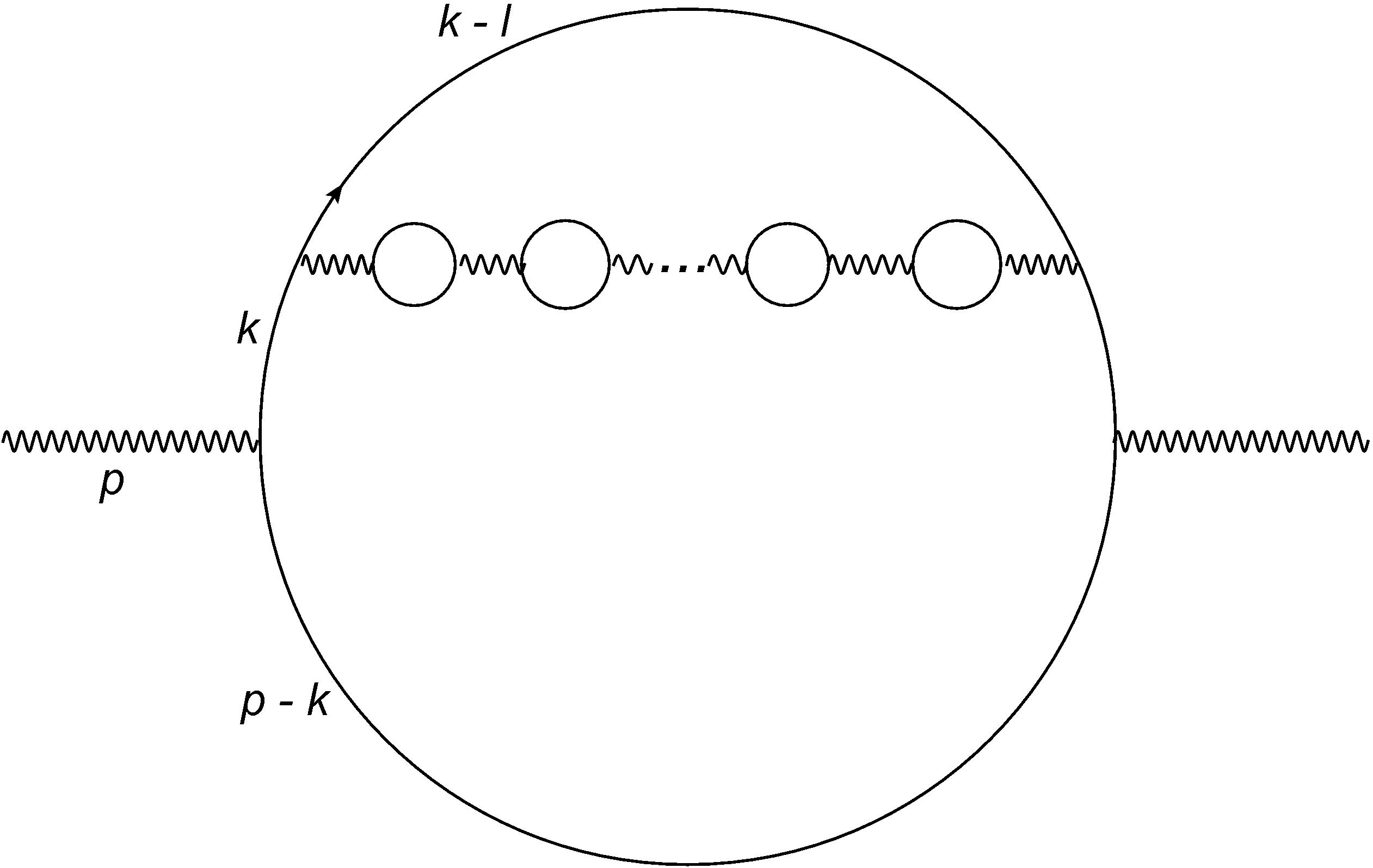}\hspace{0.05\columnwidth}
\end{minipage}
}
\subfigure[]{
\label{bubble_2}
\begin{minipage}{6cm}
\centering
\includegraphics[width=1\columnwidth]{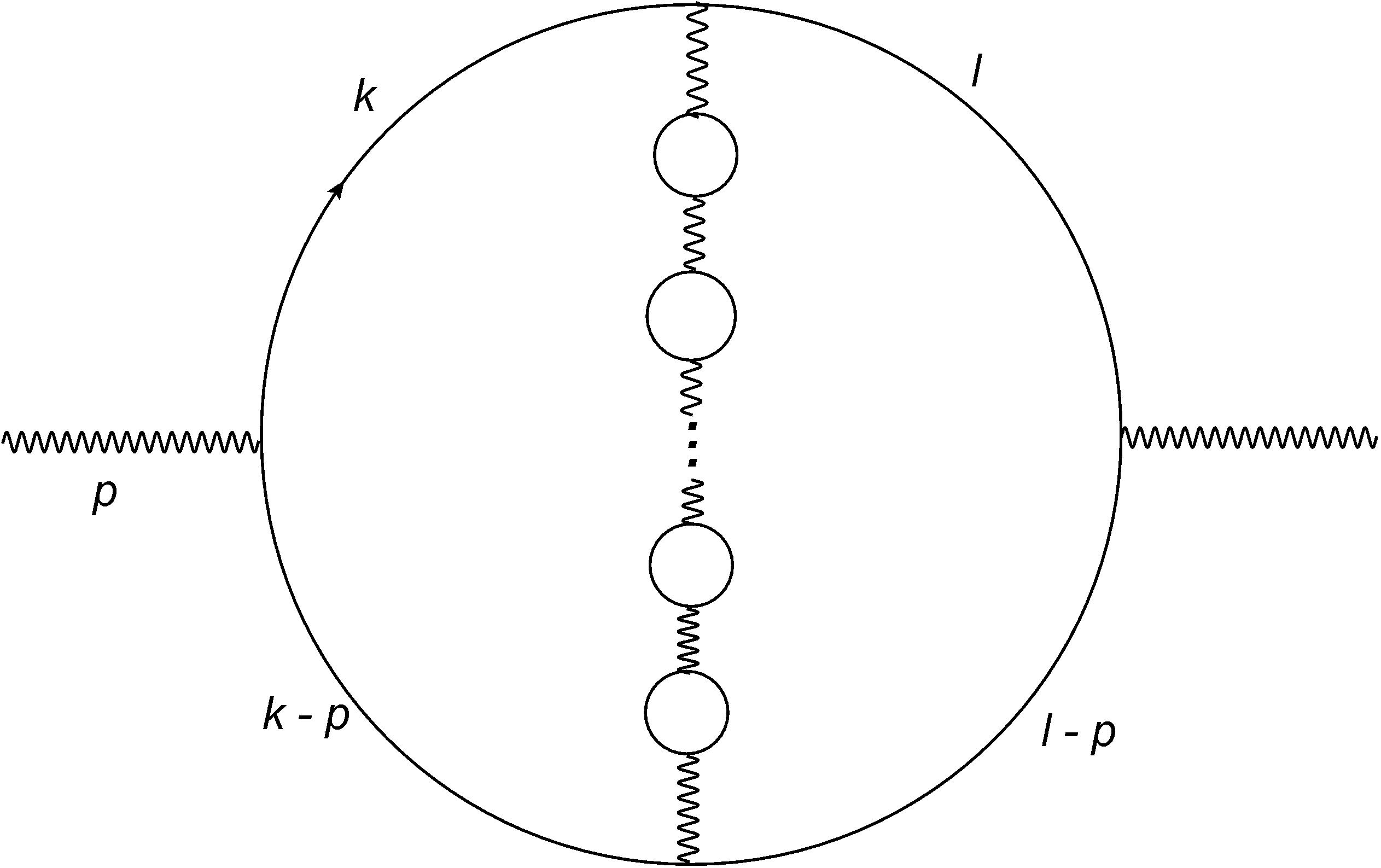}\hspace{0.05\columnwidth}
\end{minipage}
}
\subfigure[]{
\label{bubble_3}
\begin{minipage}{6cm}
\centering
\includegraphics[width=1\columnwidth]{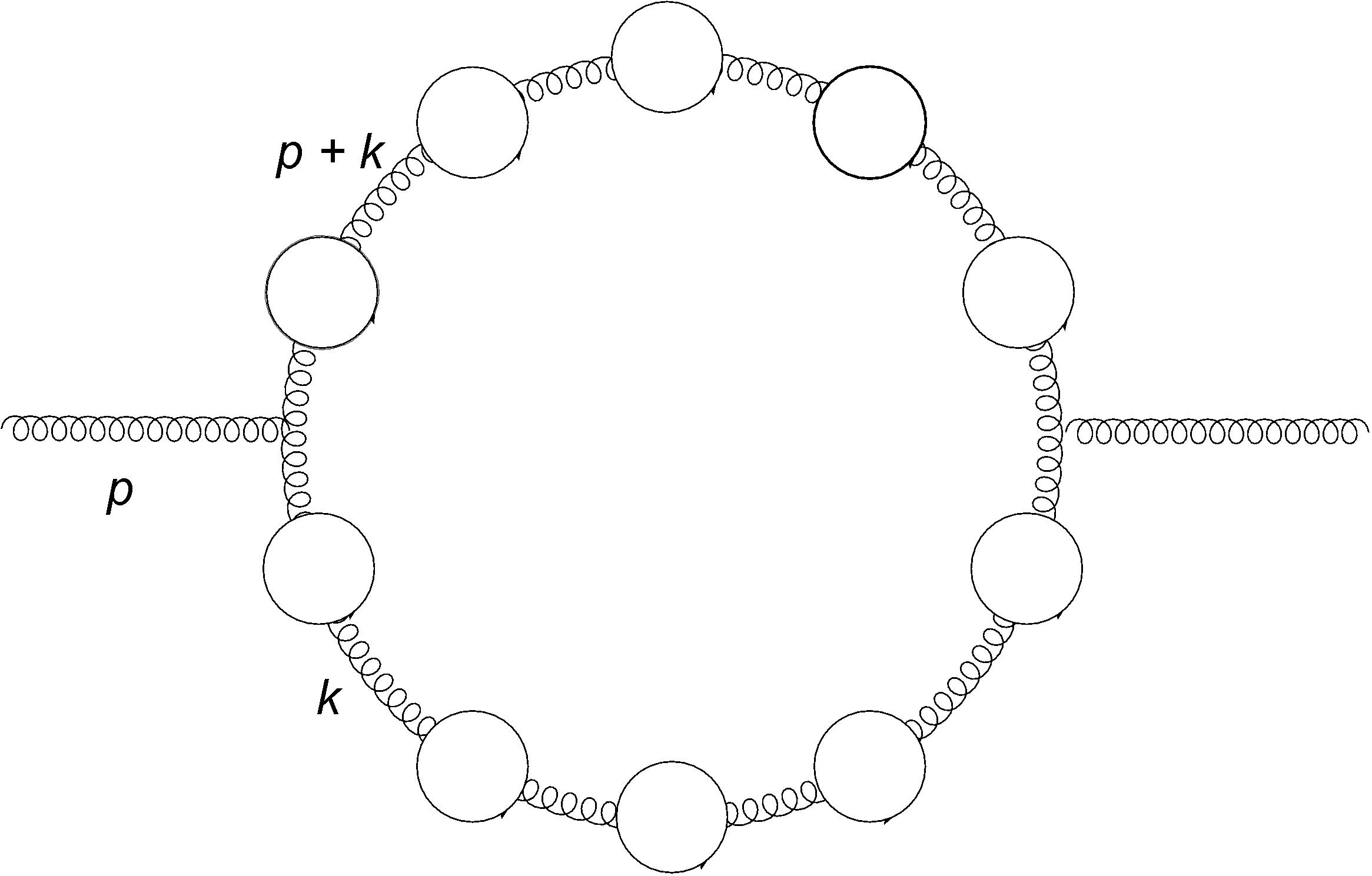}\hspace{0.05\columnwidth}
\end{minipage}
}
\subfigure[]{
\label{gauge_cancel}
\begin{minipage}{6cm}
\centering
\includegraphics[width=1\columnwidth]{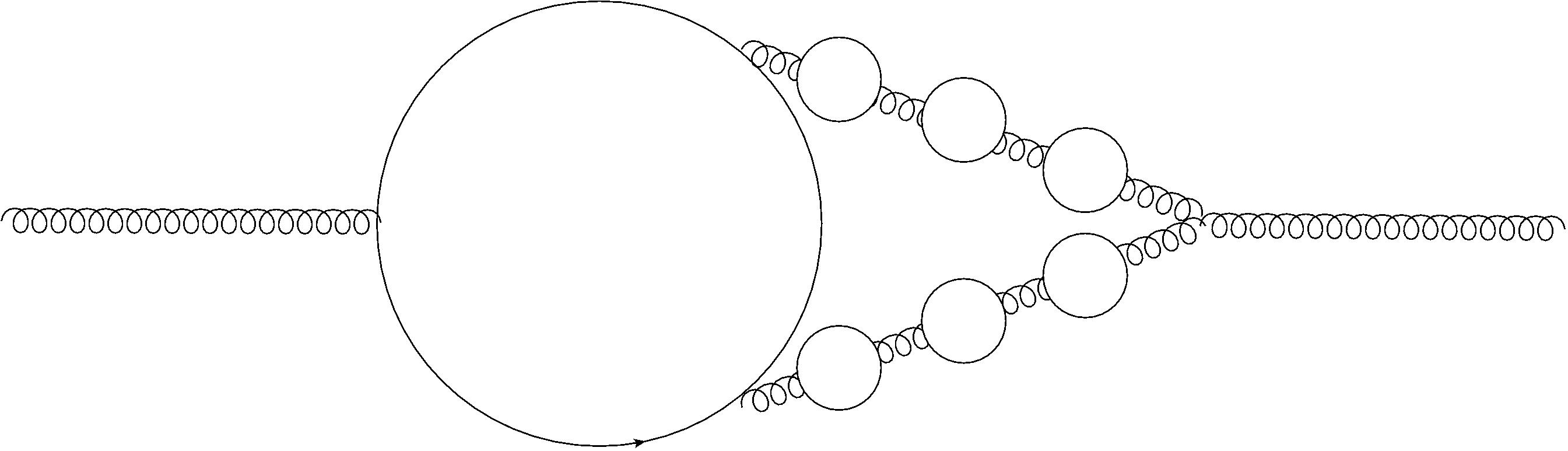}\hspace{0.05\columnwidth}
\end{minipage}
}
\caption{Feynman diagrams for gauge field renormalization at order $1/\mathcal{N}$. Diagrams (a) and (b) are present in both the Abelian and non-Abelian 2-point functions, while (c) and (d) only exist in the non-Abelian theory. }
\label{bubble_diagram}
\end{figure}

To determine the gauge beta function one has to compute the divergent part of the 2-point function. 
The leading order (LO) contribution in $1/\mathcal{N}$ is simply given by one $ \Psi $ bubble. The NLO contribution, on the other hand, is non-trivial and was computed in \cite{PalanquesMestre:1983zy,Gracey:1996he} by evaluating the diagrams shown in Fig.~\ref{bubble_diagram}. The first two  diagrams of Fig.~\ref{bubble_1} and Fig.~\ref{bubble_2} yield 
\begin{align}
\label{eq:gauge diagrams}
i\delta^{AB} p^2 \Delta^{\mu\nu}(p) \Pi^{(n)}_{P}(p)=& -( - i \tilde{g}_0)^4  \Tr \left[ T_\Psi^A T_\Psi^C T_\Psi^C T_\Psi^B \right] \mu^{2\epsilon} \int \frac{\dd^d k}{(2\pi)^d} \frac{\dd^d \ell}{(2\pi)^d} D^{(n)}_{\rho\sigma}(\ell) \nonumber \\
&\qquad \times  \Tr \left[ \gamma^{\mu}  \frac{i \slashed{k}}{k^2}  \gamma^{\rho}  \frac{i (\slashed{k} - \slashed{\ell}) }{(k - \ell)^2} \gamma^{\sigma} \frac{i \slashed{k} }{k^2} \gamma^{\nu} \frac{i(\slashed{p} - \slashed{k})}{(p-k)^2} \right], \\
i\delta^{AB} p^2 \Delta^{\mu\nu}(p) \Pi^{(n)}_{T}(p)=& -( - i \tilde{g}_0)^4 \text{Tr}\left[ T_\Psi^A T_\Psi^C T_\Psi^B T_\Psi^C \right] \mu^{2\epsilon} \int \frac{\dd^d k}{(2\pi)^d} \frac{\dd^d \ell}{(2\pi)^d} D^{(n)}_{\rho\sigma}(k-\ell) \nonumber \\
&\qquad \times \text{Tr}\left[\gamma^{\mu}  \frac{i (\slashed{k} - \slashed{p}) }{ (k- p)^2 } \gamma^{\rho} \frac{i (\slashed{\ell} - \slashed{p}) }{( \ell - p)^2 } \gamma^{\nu} \frac{i \slashed{\ell}}{\ell^2} \gamma^{\sigma}  \frac{i \slashed{k} }{k^2} \right] ,
\end{align}
where  $n\geq 0$ is the number of bubbles in the chain. These diagrams are present for all gauge groups\footnote{In the Abelian case one replaces the gauge generators with the fermion charges $ q_\Psi $.}. For the purpose of summing the contributions from all $ n $, it is useful to extract the coupling and group structure from the 2-point functions
	\begin{equation}\begin{split}
	\Pi^{(n)}_P &=  \dfrac{d(G)}{\mathcal{N}}  {K}_0^{n+2} A^{(n)}_P\ , \\
	\Pi^{(n)}_{T} &= \dfrac{d(G)}{\mathcal{N}} \left( 1 - \frac{1}{2} \frac{C_2(G)}{C_2(R_\Psi)} \right) {K}_0^{n+2}  A^{(n)}_T \ .
	\end{split}\end{equation}
Here the functions $A^{(n)}_{P,T}$ contain the loop structure of the respective diagrams and $C_2(R_\Psi)$ is the quadratic Casimir of the representation $ R_\Psi $. 

Going to the non-Abelian group we will have additional contributions from the gluon self-interactions, c.f. Fig.~\ref{bubble_3} and Fig.~\ref{gauge_cancel}. The coupling and group structure from their contribution to the 2-point function is parametrized by
	\begin{equation}
	\Pi^{(n)}_G =  \dfrac{d(G)}{\mathcal{N}} \frac{C_2(G)}{C_2(R_\Psi)}  {K}_0^{n+1}  A^{(n)}_G \ . 
	\end{equation}

\bigskip
 We now review the final results for the gauge beta functions for the Abelian and non-Abelian gauge groups.

\subsubsection{Abelian beta function}
We consider the case where the $ \Psi $ fermions are charged under a single $ \U(1) $ gauge group and determine the associated gauge beta function. In this case, we point out that the $ K $ coupling from Eq. \eqref{eq:K_coupling} reduces to $ K = g^2 q_\Psi^2 N_f /4 \pi^2 $ that agrees with earlier literature.  The resummation of the beta function was performed first by Palanques-Mestre and Pascual \cite{PalanquesMestre:1983zy}. Including both the LO and the $ 1/\mathcal{N} $ contributions to the 1PI 2-point function, they found\footnote{The function $\tilde{F}$ is related to the Mestre-Pascual result via $ \tilde{F}(n,\epsilon) = F_\mathrm{MP}(n,-\epsilon/2) $.}
	\begin{equation}
	\begin{split}
	Z_K \Pi_B &= Z_K \Pi_0(p^2) + Z_K \dfrac{{K}_0}{\mathcal{N}} \sum_{n=1}^{\infty} {K}_0^n \left( A^{(n-1)}_T  +2 A^{(n-1)}_P \right) \\
	&= - 2K \left(-\dfrac{4\pi\mu^2}{p^2} \right)^{\epsilon/2} \Gamma_0(\epsilon) + \frac{3K}{4\mathcal{N}}  \sum_{n=1}^{\infty} \left( -\frac{2K_0}{3} \right)^n \frac{1}{(n+1)\epsilon^{n}}  \tilde{F}(n+1,\epsilon) \ .
	\end{split}
	\end{equation}
Notice that we have used the fact that for Abelian gauge theory $ Z_K = Z_A $. The function $ \tilde{F} $  encodes the diagram structure, and it turns out that the beta function depends only on $ \tilde{F}(0,\epsilon) $. Using analyticity of $\tilde{F}$,  we can apply the resummation formula \eqref{eq:fourth_resum} to obtain 
	\begin{equation}\label{eq:U1_Z}
	Z_K^{(1)} = \left. Z_K \Pi_B \right|_{1/\epsilon} = -\dfrac{2K}{3} - \frac{1}{2\mathcal{N}} \int_0^{K} \dd x\, (K-x) \tilde{F} (0,\tfrac{2}{3}x)\ ,
	\end{equation}
where
	\begin{equation}
	\tilde{F}(0,x) = \dfrac{(1-x) (1-\tfrac{x}{3}) (1+\tfrac{x}{2})\Gamma(4-x) }{3 \Gamma^2(2-\tfrac{x}{2}) \Gamma(3-\tfrac{x}{2}) \Gamma(1+\tfrac{x}{2})} \ .
	\end{equation}	
Finally, applying   \eqref{eq:beta_formulas}  the gauge beta function reads
	\begin{equation}\begin{split}
	\beta_K &=  -K^2 \dfrac{\partial}{\partial K} Z_K^{(1)} = \frac{2K^2 }{3} \left[ 1 + \frac{1}{\mathcal{N}} F_1(K)  \right] 
	\end{split}\end{equation}
to NLO in $ 1/\mathcal{N} $. For later convenience we  introduced 
	\begin{equation}
	F_1(K) = \frac{3}{4} \int^{K}_0 \dd x\, \tilde{F}\left( 0, \tfrac{2}{3} x \right) \ .\label{abelian_summation}
	\end{equation}

\subsubsection{Non-Abelian beta function}
Now we turn to the case where the $ \Psi $ fermions transform under a given representation $R_\Psi$ of  a non-Abelian gauge group. In this instance, the gauge field 2-point function gets an additional contribution due to the gluon self-interaction. The NLO 2-point function is then given by 
	\begin{equation}
	\begin{split} \label{eq:non_Abelian_2point}
	\Pi_B = \Pi_0 + K_0 \frac{d(G)}{\mathcal{N}}\sum_{n=1} &\left\{ K_0^{n} \left[\left( 1 - \frac{C_2(G)}{2 C_2(R_\Psi)} \right) A^{(n-1)}_T + 2 A^{(n-1)}_P \right] \right. \\
	&\qquad \left.  + K_0^{n-1}\frac{C_2(G)}{ C_2(R_\Psi)} A_G^{(n-1)}\right\} \ .
	\end{split}
	\end{equation}
In the non-Abelian case the gauge coupling renormalization is more involved. The computation can either be performed in the background field gauge or in $ \xi $ gauge provided that for the latter one includes the vertex renormalization. This computation was originally performed in \cite{Gracey:1996he} using an alternative method and later reviewed in \cite{Holdom:2010qs} and the result reads
	\begin{equation}
	\beta_K = \dfrac{2K^2}{3} \left[1 + \dfrac{d(G)}{\mathcal{N}} H_1(K) \right] \ ,
	\label{QCDbeta}
	\end{equation}
where we have defined the functions
\begin{equation}\begin{split}\label{eq:H_1}
H_1(K) &= -\dfrac{11 C_2(G)}{4 C_2(R_\Psi)} +\frac{3}{4} \int_0^K \dd x\, \tilde{F}(0,\tfrac{2}{3}x) \tilde{G}(\tfrac{1}{3}x)\ ,\\
\tilde{G}(x) &= 1+\frac{C_2(G)}{C_2(R_{\Psi})}\frac{20-43x+32{x}^{2}-14{x}^{3}+4{x}^{4} }{ 4\left( 2\,x-1 \right)  \left(2\,x-3 \right) \left( 1-x^2 \right) }\ .
\end{split}\end{equation}
While $\tilde{F}$ is the function obtained in the $\U(1)$ case, the non-trivial part of $\tilde{G}$  stems from the gluon contribution\footnote{$\tilde{G}\left(x\right)=\frac{d(R_\Psi)}{d\left(G\right)}I_2\left(x\right)$ when comparing with the result in \cite{Holdom:2010qs}.}. Notice that $ H_1 $ reduces to $ F_1 $ in the case of an Abelian gauge group, so Eq. \eqref{QCDbeta} is valid for all simple gauge groups.

\subsection{Extension to semi-simple gauge groups}
\begin{figure}[bt]
	\centering
	{\includegraphics[width=.9\textwidth]{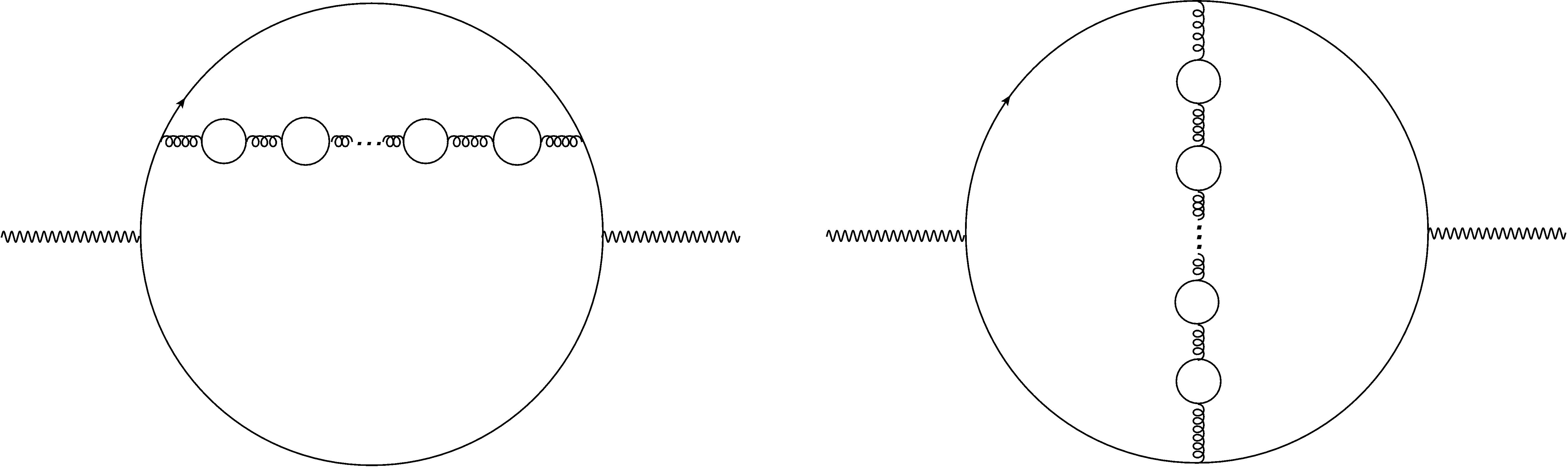}}
	\caption{Feynman diagrams for the 2-point functions giving mixed terms to the beta functions.}
	\label{bubble_semi_simple} 
\end{figure} 
Let us now generalize the result to the case where the vector-like fermions are charged under a semi-simple gauge group. To determine the mixed contribution to the gauge coupling renormalization $ Z_{K_\alpha} $ it is sufficient to consider only the mixed diagrams appearing in the gauge 2-point function. In the $ \xi $ gauges the mixed contribution to the vertex and fermion field renormalization will cancel against each other.  To determine $ Z_{A_\alpha} $ we employ equation \eqref{eq:gauge_field_renormalization} to accommodate mixed gauge contributions. Starting from one gauge field, $ \alpha $, the 2-point function contains the usual terms present in Eq. \eqref{eq:non_Abelian_2point}. These are unaffected by the presence of other gauge groups. Additionally, at NLO, it is possible to have a gauge bubble chain from a different gauge group $ \beta $ stretching across the fermion loop instead of the original $ \alpha $ chain, as shown in Fig.~\ref{bubble_semi_simple}. These are the only type of diagrams mixing the gauge groups, since a single fermion bubble cannot couple simultaneously to two different gauge groups. The new contribution to the 1PI 2-point function coming from the mixed diagrams with the group $ \beta $ is
	\begin{equation}
	\begin{split}
	Z_{A_\alpha} \Delta \Pi_{\alpha,B} &=  Z_{A_\alpha} K_{\alpha,0} \dfrac{d(G_\beta)}{\mathcal{N}} \sum_{n=1}^{\infty} K_{\beta,0}^n \left(A^{(n-1)}_T  +2 A^{(n-1)}_P \right) \\
	&=\frac{3 \,d(G_\beta)}{4\mathcal{N}} K_\alpha \sum_{n=1}^{\infty} \left( -\frac{2K_{\beta,0}}{3} \right)^n \frac{\tilde{F} (n+1,\epsilon) }{(n+1) \epsilon^{n}} \ .
	\end{split}
	\end{equation}
To renormalize $ K_{\alpha,0} $, we have used the fact that $ Z_{K_\alpha} = Z_{A_\alpha} $ at LO in $ 1/\mathcal{N} $. Once again, the $ 1/\epsilon $ pole can be extracted using the resummation formula \eqref{eq:fourth_resum}. The new contribution to the gauge coupling renormalization is obtained as
	\begin{equation}
	\Delta Z_{K_\alpha}^{(1)} = \Delta Z_{A_\alpha}^{(1)} = - \frac{d(G_\beta)}{2 \mathcal{N}} K_\alpha\int_0^{K_\beta} \dd x\, \left(1-\dfrac{x}{K_\beta}\right) \tilde{F} (0, \tfrac{2}{3}x) \ .
	\end{equation} 
The mixed contributions to the beta function read
	\begin{equation}
	\begin{split}
	\Delta \beta_{K_\alpha} &= - K_\alpha \left( K_\alpha \frac{\partial}{\partial K_\alpha} + K_\beta  \frac{\partial}{\partial K_\beta} \right) \Delta Z_{K_\alpha}^{(1)}=  \frac{ d(G_\beta)}{ 2 \mathcal{N}} K_\alpha^2 \int_{0}^{K_\beta} \dd x\, \tilde{F}(0, \tfrac{2}{3}x)\ .
	\end{split} 
	\end{equation}
Taking into account the mixed contributions coming from all the different gauge groups to each beta function we find 
	\begin{equation} \label{eq:largeN_gauge_beta}
	\beta_{K_\alpha} = \dfrac{2K^2_\alpha}{3} \left[1 + \dfrac{d(G_\alpha)}{\mathcal{N}} H_1^{(\alpha)}(K_\alpha) + \dfrac{1}{\mathcal{N}} \sum_{\beta \neq \alpha} d(G_\beta) F_1(K_\beta) \right].
	\end{equation}
Note that the $ H_1 $ functions are dependent on the specific gauge group and fermion representation as evident from Eq. \eqref{eq:H_1} (hence the superscript). In the case of an Abelian group $ H_1 $ reduces  to $ F_1 $.  
 
A test of our results consists in checking that when re-expanding the beta functions  given in Eq.~\eqref{eq:largeN_gauge_beta}  as functions of the couplings the coefficients agree with the state-of-the art  3-loop perturbative computation \cite{Mihaila:2014caa} which for $G= G_\alpha \times G_\beta$ reads

\begin{equation}
 \begin{split}
 \beta_{K_\alpha}^{\rm 3-Loop}&=\dfrac{2K^2_\alpha}{3}\Bigg[1+\frac{1}{\mathcal{N}}\left(\frac{K_\alpha\left(5C_2\left(G_\alpha\right) +3C_2(R_\Psi^\alpha) \right)}{4S_2 \left( R_\Psi^\alpha \right) d(R_\Psi^\alpha)^{-1}} -\frac{ K_\alpha^2 \left(79 C_2\left(G_\alpha\right)+66C_2(R_\Psi^\alpha)\right)}{288S_2(R_\Psi^\alpha)d(R_\Psi^\alpha)^{-1}}\right)\\
 &+\frac{1}{\mathcal{N}}\left(\frac{3K_\beta C_2(R_\Psi^\beta)}{4S_2(R_\Psi^\beta)d(R_\Psi^\beta)^{-1}}-\frac{11K_\beta^2 C_2(R_\Psi^\beta)}{48S(R_\Psi^\beta)d(R_\Psi^\beta)^{-1}}\right)-\frac{1}{\mathcal{N}^2}\Bigg(\frac{17K_\alpha C_2(G_\alpha)^2}{3S_2(R_\Psi^\alpha)^2d(R_\Psi^\alpha)^{-2}}\\
&+K_\alpha^2\frac{1415C_2(G_\alpha)^2+615C_2\left(G_\alpha\right)C_2(R_\Psi^\alpha)-288C_2(R_\Psi^\alpha)^2}{288S_2(R_\Psi^\alpha)^2d(R_\Psi^\alpha)^{-2}}+\cdots\Bigg)\\
&-\frac{1}{\mathcal{N}^3}\Bigg(K_\alpha^2\frac{2857C_2(R_\Psi^\alpha)^3}{288S_2(R_\Psi^\alpha)^3d(R_\Psi^\alpha)^{-3}}+\cdots\Bigg)\Bigg]\,.\label{RG_Gauge_3loop}
 \end{split}
 \end{equation}

\noindent


\noindent
It is straightforward to check that the leading $1/\mathcal{N}$ terms agree with the corresponding terms in Eq.~\eqref{eq:largeN_gauge_beta}. 

\bigskip
In the derivation of Eq.~\eqref{eq:largeN_gauge_beta} we have assumed that the gauge group under which $ \Psi $ is charged contains at most one $ \U(1) $. If that were not the case, it would be possible for the fermion bubbles to couple to two different Abelian groups simultaneously. This would give a new class of diagrams, where the bubble chains would alternate between the two groups. In such a case one would also have to take into account kinetic mixing between the two gauge groups.  This has not be considered here.

\subsection{Safe phase diagrams}
To conclude this section, we investigate the short distance fate of gauge-fermion theories  at large number of matter fields. Here asymptotic freedom is lost and unless an interacting UV fixed point emerges, the underlying theory can be viewed, at best,  as an effective low-energy description of physical phenomena.  In this regime asymptotic safety  is  dynamically achieved due to the collective effect of the many fermions present in the theory. This is reflected in the emergence of a non-trivial zero of the  beta functions at NLO in $ 1/\mathcal{N} $~\cite{Antipin:2017ebo}.

\subsubsection{Safe QCD} 
For single gauge groups, resembling QCD with many flavors, asymptotic safety is indeed a possibility \cite{Antipin:2017ebo}.   To elucidate this point while making this work self-contained, we briefly summarize here the salient points of how a UV fixed point emerges. To make our point clear, we consider an $\SU(N)$ gauge group with $N_f$ flavors transforming according to the fundamental representation. From Eq.~\eqref{QCDbeta}, one shows that there is a fixed point a $K^{\ast} = 3$ up to exponentially vanishing corrections \cite{Antipin:2017ebo}.  This occurs because the $\tilde{G}(x)$ function in \eqref{eq:H_1} has a pole in the integrand at $x=1$ ($K=3$), corresponding to a logarithmic singularity in the $H_1(K)$ function.  The beta function will therefore  have a UV fixed point at $K^*\approx 3$  to leading order in $1/\mathcal{N}$, which is obtained from the condition $1+d(G)H_1(K)/\mathcal{N}=0$.  The new conformal window for these theories as function of number of flavors and colors extends the original infrared (IR) conformal window to also contain the asymptotically safe scenario~\cite{Antipin:2017ebo}.   It is worth mentioning that to provide a rough estimate of the lower boundary of the asymptotically safe window, one can use the stability of the $1/\mathcal{N}$ expansion \cite{Holdom:2010qs,Antipin:2017ebo} by estimating when the $1/\mathcal{N}^2$ and higher corrections become relevant.  
 
%

\subsubsection{Safe Semi-simple Gauge Groups}
\begin{figure}[t]
\centering
\subfigure[$\SU(N_1)\times \SU(N_2)$]{
\label{Phase2}
\begin{minipage}{7cm}
\centering
\includegraphics[width=1\columnwidth]{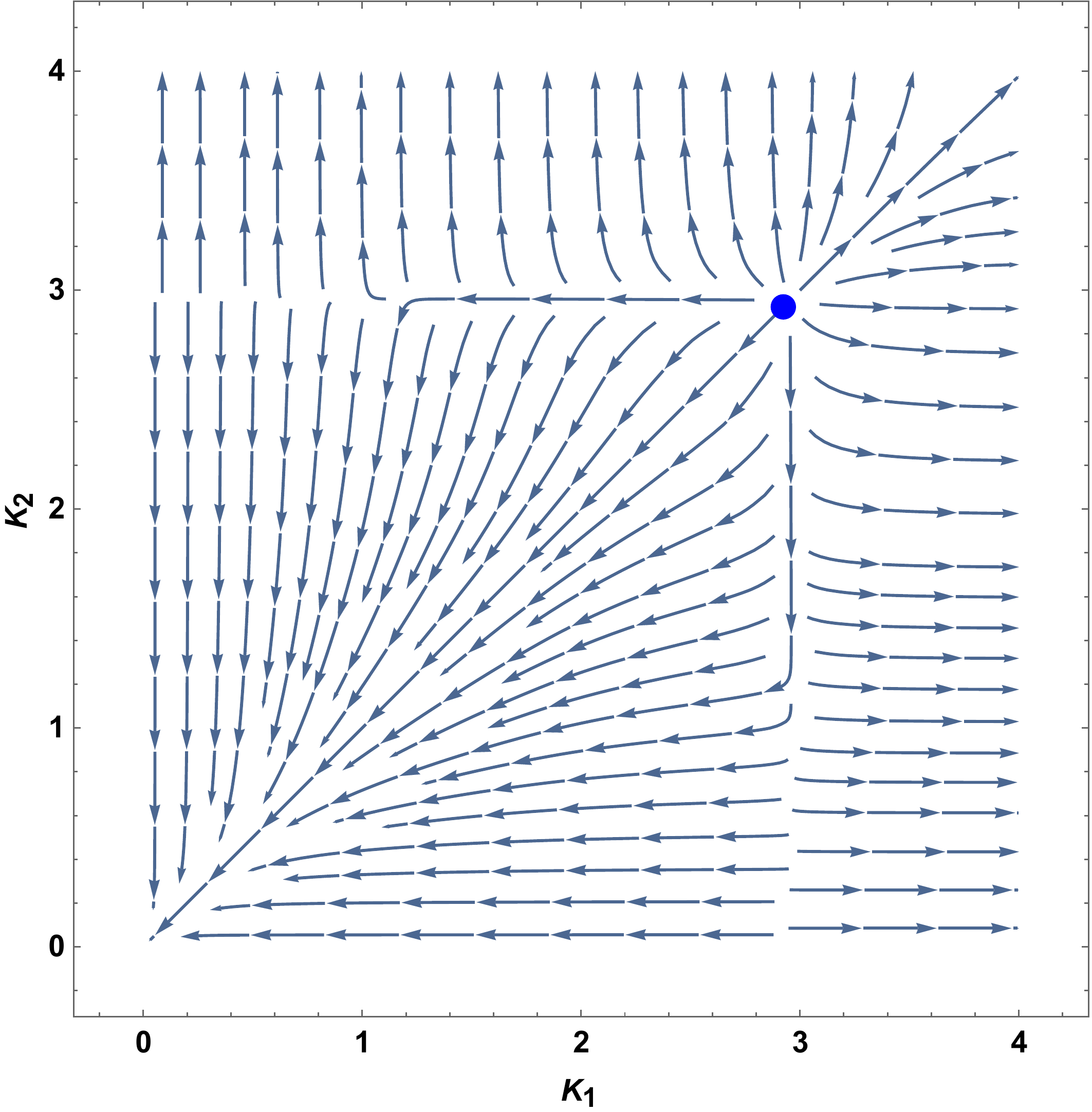}\hspace{0.05\columnwidth}
\end{minipage}
}
\subfigure[$\U(1) \times \SU(N)$]{
\label{Phase1}
\begin{minipage}{7cm}
\centering
\includegraphics[width=1\columnwidth]{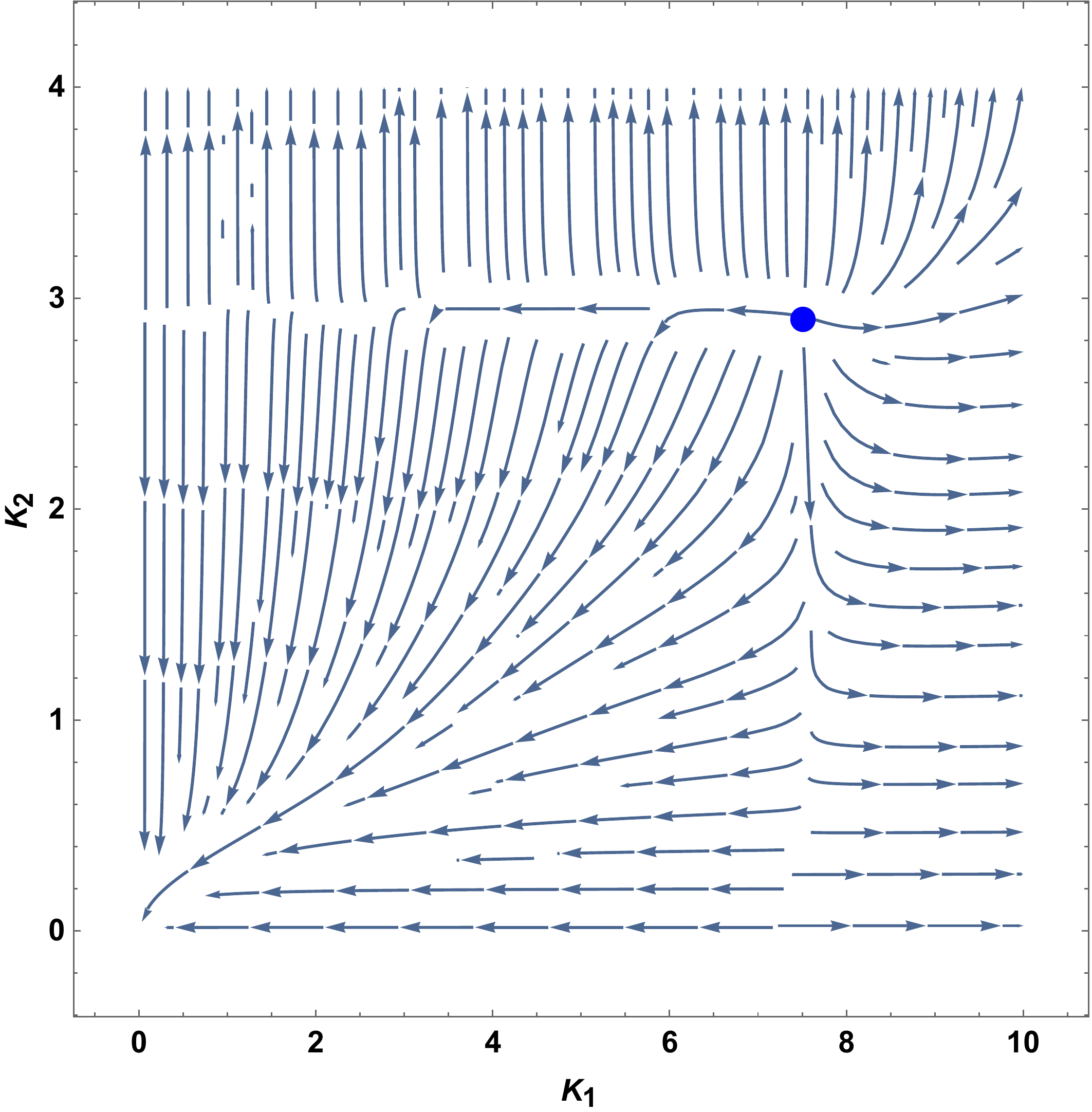}\hspace{0.05\columnwidth}
\end{minipage}
}
\caption{Phase diagrams of semi-simple gauge theories consisting of two non-Abelian groups (left) and an Abelian and a non-Abelian group (right).}
\end{figure}

 We now  investigate the semi-simple case   starting with the   $G = \SU(N_1) \times \SU(N_2)$ example.  The associated phase diagram refers to the RG-flow  plotted in the plane of the two gauge couplings, $K_1$ and $K_2$, and it is presented in Fig.~\ref{Phase2}. The UV interacting fixed point, repulsive in all directions,  occurs for $K_1^\ast = K_2^\ast =3$ (the blue-dot) up to exponentially small corrections. Two more interacting fixed points occur for $(K_1^\ast =3, K_2^\ast =0)$ and  $(K_1^\ast =0, K_2^\ast =3)$ corresponding to the fixed points of each single gauge group. Finally we have the Gaussian IR fixed point at the origin of the coupling space. This analysis complements the perturbative analysis for semi-simple gauge groups investigated first in \cite{Esbensen:2015cjw}. We therefore discover that there is a UV complete fixed point for semi-simple gauge theories with a two-dimensional critical surface.

 The phase diagram for the semi-simple group $G = \U(1) \times \SU(N_2)$ is presented in Fig.~\ref{Phase1}. It is structurally identical to the $ \SU(N_1) \times \SU(N_2)$ case above with the difference that the UV fixed point for the $\U(1)$ gauge couplings occurs at $K_1^\ast=15/2$.

One can derive a rough estimate of the asymptotically safe conformal window for the semi-simple gauge group as well. We use again the stability of the $1/\mathcal{N}$ expansion by estimating numerically the size of the known $1/\mathcal{N}^2$ and $1/\mathcal{N}^3$ corrections from  Eq.~\eqref{RG_Gauge_3loop}. 
We expect it to be wider than safe QCD because the effective number of flavors $\mathcal{N} = N_f d(R_\Psi^{1})d(R_\Psi^{2}) $ is  larger. 
%



\section{Yukawa and self-coupling beta functions}\label{sec:SelfishYukawa}
We now review and further elucidate the computation of the RG-functions of the Yukawa \cite{Kowalska:2017pkt} and quartic couplings \cite{Pelaggi:2017abg} of the model \eqref{eq:L_model} in  the presence of a large number of vector-like fermions.  Finally, the results of the running of the quartic coupling are extended to the case where the $ \Psi $ fermions transform under a semi-simple gauge group. We are interested in the case in which $y$ and $\lambda$ scale with $\mathcal{N}$ as $ \lambda \sim y^2 \sim 1/{\mathcal{N}}$. This is the region for which a UV fixed point can appear due to the interplay between the large $\mathcal{N}$ gauge contribution and leading  corrections stemming from the Yukawa and scalar self-coupling. With this scaling of the couplings it is  sufficient to consider  the 1-loop contributions from the Yukawa and quartic coupling to their beta functions. The counting will then ensure that higher loops will give corrections that are higher order in $ 1/\mathcal{N} $. 

The leading $1/{\mathcal{N}}$ contribution stemming from the $ \Psi $ fermions is obtained by dressing gauge lines with their bubbles. As in section \ref{sec:gauge_fermion}, these diagrams can be resummed and the $ 1/\epsilon $ pole extracted in a closed from. We will first discuss the new contribution to the fermion and scalar self energies before moving to compute the vertex corrections.  We shall see that it is straightforward to generalize the results to the semi-simple case except for the quartic self-coupling.

\subsection{Fermion self-energy}
To compute the Yukawa beta function we need first to compute the gauge correction to the fermion self-energy to LO in $ 1/\mathcal{N} $ for the $ \chi, \xi $ fermions.

\subsubsection{Abelian case}
\begin{figure}[t]
	\centering
	\includegraphics[width=0.5\columnwidth]{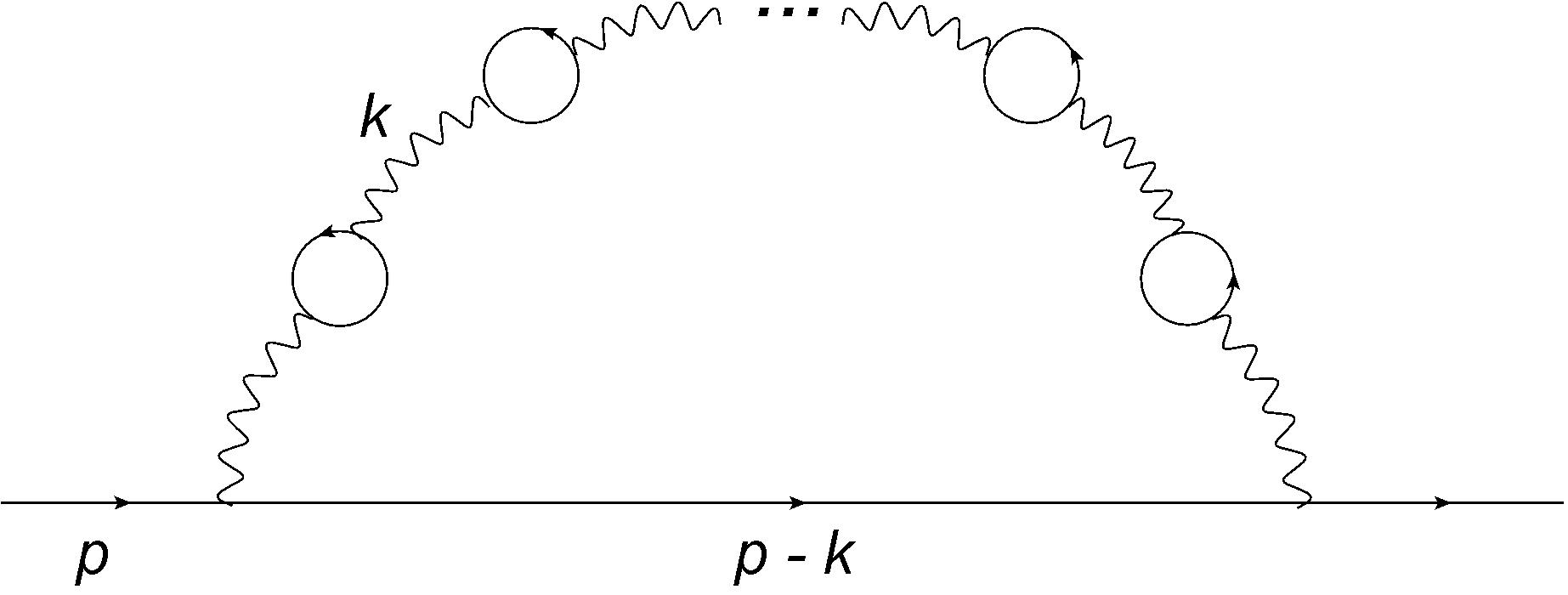}\hspace{0.05\columnwidth}
	\caption{LO gauge contribution to the fermion self-energy.}
	\label{fermion_self_energy}
\end{figure}

We start with the Abelian case and then extend the result to the non-Abelian one.  At this order in $ 1/\mathcal{N} $ the relevant diagram is shown in Fig.~\ref{fermion_self_energy}. 
For  the $ \chi $ (identically for $\xi$) self-energy, the defining integral for the $ n $-bubble diagram is
	\begin{equation}\label{eq:fermion_se_diagram}
	-i \Sigma_\chi^{(n)}(p) = ( i q_\chi \tilde{g}_0)^2 \mu^{\epsilon} \int \frac{\dd^d k}{(2\pi)^d} \bar{\sigma}^{\mu} \frac{i\sigma \cdot (p-k)}{(p-k)^2} \bar{\sigma}^{\nu} D^{(n)}_{\nu\mu}(k) \ .
	\end{equation} 
 The integral is known, and the diagram evaluates to 
	\begin{equation}
	-i \Sigma^{(n)}_\chi (p) = -\frac{3i}{8 \mathcal{N}} \frac{q^2_\chi}{q_\Psi^2}  \bar{\sigma} \cdot p \left( -\frac{2 K_0}{3} \right)^{n+1} 3^n \Gamma_0^n(\epsilon) \Gamma_\psi(n,\epsilon) \left( - \frac{4 \pi \mu^2}{p^2} \right)^{(n+1)\epsilon/2}, 
	\end{equation}
where we defined
	\begin{equation}
	\Gamma_\psi(n,\epsilon) = \dfrac{n (3 - \epsilon)}{n+1} \dfrac{ \Gamma(2 - \tfrac{\epsilon }{2} ) \Gamma(1 + \tfrac{n+1}{2} \epsilon ) \Gamma(1 - \frac{n+1}{2} \epsilon)}{\Gamma(2+ \tfrac{n}{2}\epsilon ) \Gamma(3 - \frac{n+2}{2} \epsilon)} \ .
	\end{equation}
Summing over all bubbles to obtain the total gauge contribution to the self-energy at $ 1/\mathcal{N} $ and shifting the sum from $n \rightarrow n-1 $, we obtain 
	\begin{equation}
	\frac{\dd \Sigma_\chi}{\dd \bar{\sigma} \cdot p} = - \frac{9}{16 \mathcal{N}} \frac{q^2_\chi}{q_\Psi^2}  \sum_{n=1}^{\infty} \left(- \frac{2 K_0}{3} \right)^n \dfrac{1}{n \epsilon^n} H_\psi(n, \epsilon) \ ,
	\end{equation}
where
	\begin{equation}
	H_\psi(n,\epsilon) = - \dfrac{2}{3} \left(- \frac{4 \pi \mu^2}{p^2} \right)^{n\epsilon/2} [3\epsilon \, \Gamma_0(\epsilon)]^{n-1} (1-n\epsilon) n \epsilon\,  \Gamma_\psi(n-1,\epsilon) \ .
	\end{equation}
The contribution to the RG-function stems from the  $ 1/\epsilon $ pole which is extracted using the resummation formula \eqref{eq:first_resum} and yields  
	\begin{equation}
	Z_{\chi (\xi) }^{(1)} = \pole{\frac{\dd \Sigma_{\chi(\xi)}}{\dd \bar{\sigma} \cdot p}}  = \frac{3}{8 \mathcal{N}} \frac{q_{\chi (\xi)}^2}{q_\Psi^2} \int_{0}^{K} \dd x \, H_\psi(0,\tfrac{2}{3}x). 
	\end{equation}
To arrive at the above relation between $Z_\chi$ and the 2-point function  we used the fact that $Z_\chi = 1 + \mathcal{O}(1/\mathcal{N}) $. For the reader's convenience we also give the expression for $H_\psi$ 
	\begin{equation}
	H_{\psi}(0,x) = \dfrac{x(1-\tfrac{x}{3}) \Gamma(4-x)}{3 \Gamma^2(2-x) \Gamma(3-\tfrac{x}{2} ) \Gamma(1+\tfrac{x}{2})} \ .
	\label{sum_fermion}
	\end{equation}

\subsubsection{Non-Abelian case}
The result for the non-Abelian gauge group case is obtained by replacing 	
	\begin{equation}
	(q_\chi \tilde{g}_0)^2 \longrightarrow \tilde{g}_0^2 (T_\chi^{A} T_\chi^{A})\udindices{i}{j} = \dfrac{4\pi^2 d(R_\Psi) K_0}{\mathcal{N}\, S_2(R_\Psi)} C_2(R_\chi) \, \delta\udindices{i}{j} 
	\end{equation}
in the $ n $-bubble self-energy \eqref{eq:fermion_se_diagram}. The rest of the computation follows the Abelian case  yielding the field-strength renormalization  
	\begin{equation}
	Z_{\chi (\xi)}^{(1)} =  \frac{3d(R_\Psi)}{8 \mathcal{N}\,S_2(R_\Psi)} C_2(R_{\chi (\xi)}) \int_{0}^{K} \dd x \, H_\psi(0,\tfrac{2}{3}x) \ . 
	\end{equation}

\subsection{Scalar self-energy}
\begin{figure}[t]
	\centering
	\includegraphics[width=0.5\columnwidth]{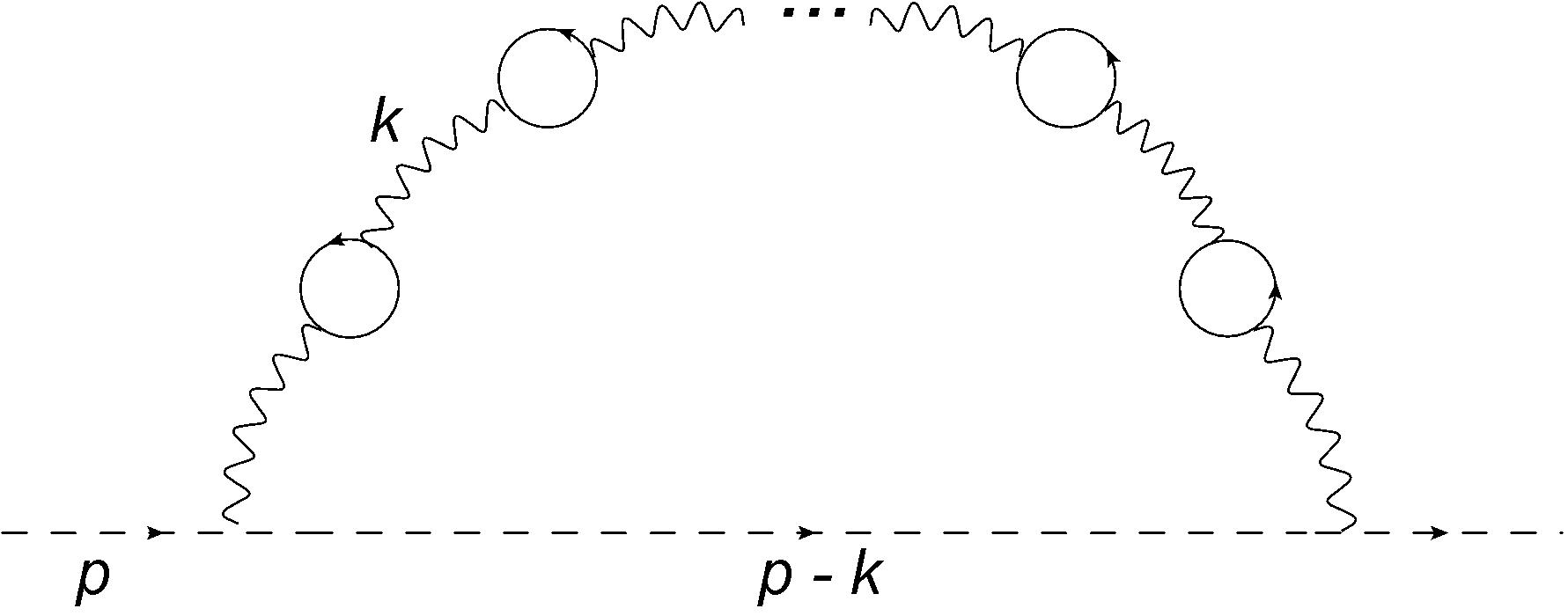}\hspace{0.05\columnwidth}
	\caption{LO gauge contribution to the scalar self-energy.}
	\label{g_square}
\end{figure}
We proceed to determine the  correction to the scalar self-energy at LO in $1/\mathcal{N}$. This is a necessary step towards the full computation of the Yukawa and quartic self-coupling.  
\subsubsection{Abelian case}
  Here the  diagrams that contribute contain a chain of $ n $ fermion bubbles as shown in Fig.~\ref{g_square}. Analytically 
	\begin{equation} \label{eq:scalar_se_diagram}
	-iS^{(n)}(p^2) = (i\tilde{g}_0 q_\phi)^2 \mu^{\epsilon}\int \dfrac{\dd^d k}{(2\pi)^d} \dfrac{i}{(p-k)^2} (2p-k)^\mu D^{(n)}_{\mu\nu}(k) (2p-k)^\nu  \ .\\
	\end{equation}
The integral  yields the $ n $-bubble contribution 
	\begin{equation}
	-iS^{(n)}(p^2) = -i \dfrac{K_0 q_\phi^2}{\mathcal{N} q_\Psi^2} (-2K_0)^{n} p^2 \left(- \dfrac{4\pi \mu^2}{p^2}\right)^{(n+1) \epsilon/2} (3 - \epsilon) \Gamma_0^n(\epsilon) \, \Gamma_\phi(n,\epsilon)\ , 
	\end{equation}
where we have defined the function
	\begin{equation}
	\Gamma_\phi(n,\epsilon) = \dfrac{ \Gamma(1-\tfrac{n+1}{2}\epsilon) \Gamma(2-\tfrac{\epsilon}{2}) \Gamma(\tfrac{n+1}{2}\epsilon) }{2 \, \Gamma(2+\tfrac{n}{2}\epsilon) \Gamma(3- \tfrac{n+2}{2}\epsilon)}\ .
	\end{equation}
Summing over $ S^{(n)}(p^2) $ and shifting the summation from $ n $ to $ n-1 $, one can rewrite the derivative with respect to $ p^2 $  in the form  
	\begin{equation}\label{eq:dS_scalar_H}
	\dfrac{\dd}{\dd p^2}S(p^2)  = - \dfrac{9 q_\phi^2}{4\mathcal{N} q_\Psi^2} \sum_{n=1}^{\infty} \left(-\dfrac{2 K_0}{3}\right)^n \dfrac{1}{n\epsilon^n} H_\phi(n,\epsilon) \ .
	\end{equation}
Here we defined
	\begin{equation}\label{eq:scalar_H}
	H_\phi(n,\epsilon) =  4 \left(- \dfrac{4\pi \mu^2}{p^2}\right)^{n \epsilon/2} (1 - \tfrac{n}{2}\epsilon ) (1 - \tfrac{\epsilon}{3} ) \left[3\epsilon \, \Gamma_0(\epsilon)\right]^{n-1} \Gamma_\phi(n-1,\epsilon) \ . 
	\end{equation}
The simple $ \epsilon $ pole of interest for the RG-function,  is determined using   \eqref{eq:first_resum} and it yields	
\begin{equation}
	Z_\phi^{(1)} = \pole{\dfrac{\dd}{\dd p^2} S(p^2) } = \dfrac{3 q_\phi^2}{2 \mathcal{N} q_\Psi^2} \int_{0}^{K}\dd x\, H_\phi(0, \tfrac{2}{3}x) \ ,
	\end{equation}
where	\begin{equation}
	H_\phi(0, x) = \dfrac{(1 - \tfrac{x}{3}) \Gamma(4 - x) }{3 \Gamma^2(2-\tfrac{x}{2}) \Gamma(3- \tfrac{x}{2}) \Gamma(1 + \tfrac{x}{2}) } \ .
	\label{sum_scalar}
	\end{equation}

\subsubsection{Non-Abelian case}
To determine the scalar self-energy for the non-Abelian case one replaces the $\U(1)$ charges in \eqref{eq:scalar_se_diagram}  as follows 
	\begin{equation}
	(q_\phi \tilde{g}_0)^2 \longrightarrow \tilde{g}_0^2 (T_\phi^{A} T_\phi^{A})\udindices{a}{b} = \dfrac{4\pi^2 d(R_\Psi) K_0}{\mathcal{N} S_2(R_\Psi)} C_2(R_\phi)\, \delta\udindices{a}{b} \ . 
	\end{equation}
The rest of the computation is identical to the Abelian case and yields 
	\begin{equation}
	Z^{(1)}_\phi = \dfrac{3 \,d(R_\Psi)}{2 \mathcal{N}\, S_2(R_{\Psi})} C_2(R_{\phi}) \int_{0}^{K}\dd x\, H_\phi(0, \tfrac{2}{3}x) \ .
	\end{equation}

\subsection{Yukawa vertex}
\begin{figure}[t]
	\centering
	\includegraphics[width=0.4\columnwidth]{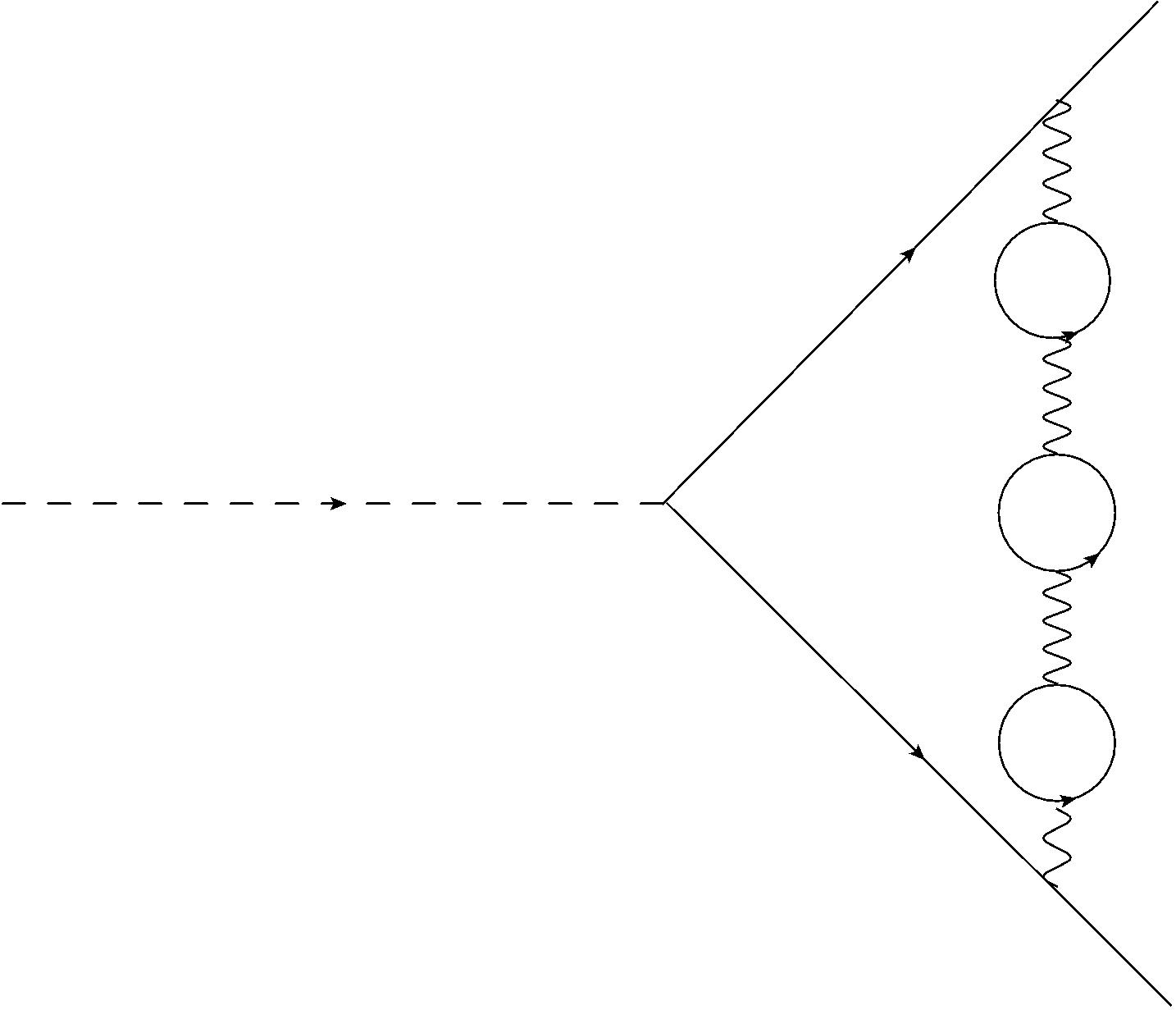}\hspace{0.05\columnwidth}
	\caption{Contributions to the Yukawa vertex at order $ 1/\mathcal{N} $.}
	\label{Yukawa_Vertex}
\end{figure}

The only vertex diagram that contributes to the Yukawa beta function in the Landau gauge is shown in Fig.~\ref{Yukawa_Vertex}. The other diagrams vanish trivially in this gauge when the external momenta are set to zero. 

\subsubsection{Abelian case}
With vanishing external momenta,  the analytic expression representing the diagrams  contributing to the Yukawa coupling with $ n $ bubbles on the gauge line is  

	\begin{equation} \label{eq:y_diagram}
	iY^{(n)} =(-i y) (-iq_\chi \tilde{g}_0)  (i q_\xi\tilde{g}_0) \mu^{\epsilon} \int\dfrac{\dd^d k}{(2\pi)^d} \sigma^\mu \dfrac{i \bar{\sigma} \cdot k }{k^2 -m^2} \dfrac{i\sigma \cdot k}{k^2 -m^2} \bar{\sigma}^\nu D_{\mu\nu}^{(n)}(k) \ .
	\end{equation}
A common mass has been added to the fermion propagators as an IR regulator, as it does not influence the divergent part of the diagram. One finds that
	\begin{equation}
	iY^{(n)} = -\dfrac{i 3\, y}{8 \mathcal{N}} \dfrac{q_{\chi} q_{\xi}}{q_\Psi^2} (-2K_0)^{n+1} \Gamma_0(\epsilon)^n \left(\dfrac{4\pi \mu^2}{m^2} \right)^{(n+1)\epsilon/2} (1-\tfrac{\epsilon}{3}) \dfrac{\Gamma(2 - \tfrac{n+1}{2}\epsilon) \Gamma(\tfrac{n+1}{2} \epsilon) }{\Gamma(2 - \tfrac{\epsilon}{2})} \ .
	\end{equation}
Next we sum over every number of fermion bubbles to find the full $ 1 / \mathcal{N} $ gauge contribution to the vertex, and cast the expression in a suitable form, yielding 
	\begin{equation}
	iY = -i y \dfrac{ 9}{ 4\mathcal{N}} \dfrac{q_\chi q_\xi}{q_\Psi^2} \sum_{n=1}^{\infty} \left(- \dfrac{2K_0}{3} \right)^{n} \dfrac{1}{n\epsilon^n} H_y(n,\epsilon)\ ,
	\end{equation}
where  
	\begin{equation}
	H_y(n,\epsilon) = \left(\dfrac{4\pi \mu^2}{m^2} \right)^{n\epsilon /2} \left[3\epsilon \, \Gamma_0(\epsilon)\right]^{n-1} (1 - \tfrac{\epsilon}{3}) \dfrac{\Gamma(2 - \tfrac{n}{2}\epsilon) \Gamma(1+ \tfrac{n}{2} \epsilon) }{\Gamma(2 - \tfrac{\epsilon}{2})}\ .
	\end{equation}
Following the usual procedure the $ 1/ \epsilon $ pole can be extracted in a closed form using  \eqref{eq:first_resum}, as $ H_y $ is sufficiently regular. The counter term for the Yukawa coupling is then extracted via 
	\begin{equation}
	\delta y^{(1)} = \left. Y \right|_{1/\epsilon} = y \dfrac{3 }{2 \mathcal{N} } \dfrac{q_\chi q_\xi}{q_\Psi^2} \int_{0}^{K}\dd x\, H_y(0,\tfrac{2}{3} x) 
	\end{equation}
with
	\begin{equation}
	H_y(0,x) = \frac{\left(1-\frac{x}{3}\right) \Gamma(4-x)}{6 \Gamma^3(2- \tfrac{x}{2}) \Gamma(1 + \tfrac{x}{2})}\ .
	\label{sum_yukawa}
	\end{equation}

\subsubsection{Non-Abelian case}
The previous result can be extended to the non-Abelian case provided that we use
	\begin{equation}
	y q_\chi q_\xi \tilde{g}_0^2 \longrightarrow y_{akl} \, \tilde{g}_0^2 (T_\chi^{A})\udindices{k}{i} (T_\xi^{A})\udindices{l}{j}  = y_{akl} \,  (T_\chi^{A})\udindices{k}{i} (T_\xi^{A})\udindices{l}{j} \dfrac{4\pi^2 d(R_\Psi) K_0}{\mathcal{N}\, S_2(R_\Psi)} 
	\end{equation}
in the Abelian expression  \eqref{eq:y_diagram}. Employing the identity  
	\begin{equation}
	y_{akl} (T^{A}_{\chi})\udindices{k}{i} (T^{A}_{\xi})\udindices{l}{j}  = -y_{aij} \dfrac{C_2(R_{\chi}) + C_2(R_{\xi}) - C_2(R_{\phi})}{2} \ ,
	\end{equation}
 we arrive at 
 	\begin{equation}
	\delta y^{(1)}_{aij} = -y_{aij} \dfrac{3 d(R_\Psi)}{2\mathcal{N}} \frac{C_2(R_\chi) + C_2(R_\xi) - C_2(R_{\phi})}{2 S_2(R_{\Psi}) } \int_{0}^{K}\dd x\, H_y(0,\tfrac{2}{3} x) \ .
	\end{equation}

\subsection{Quartic vertex}
\begin{figure}[t]
	\centering
	\includegraphics[width=0.5\columnwidth]{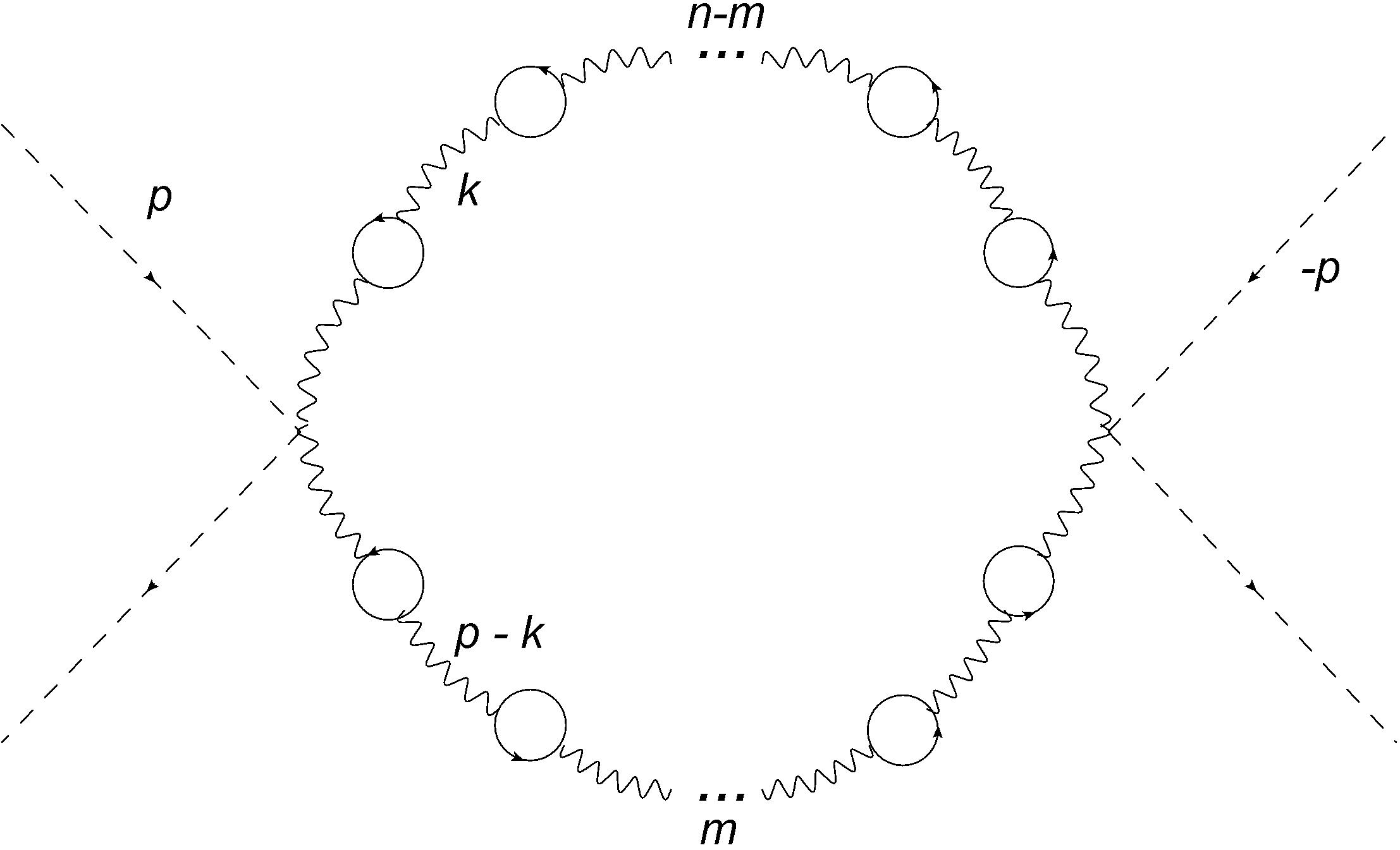}\hspace{0.05\columnwidth}
	\caption{$g^4$ vertex contribution}
	\label{fig:g_4}
\end{figure}

\begin{figure}[t]
	\centering
	\includegraphics[width=0.6\columnwidth]{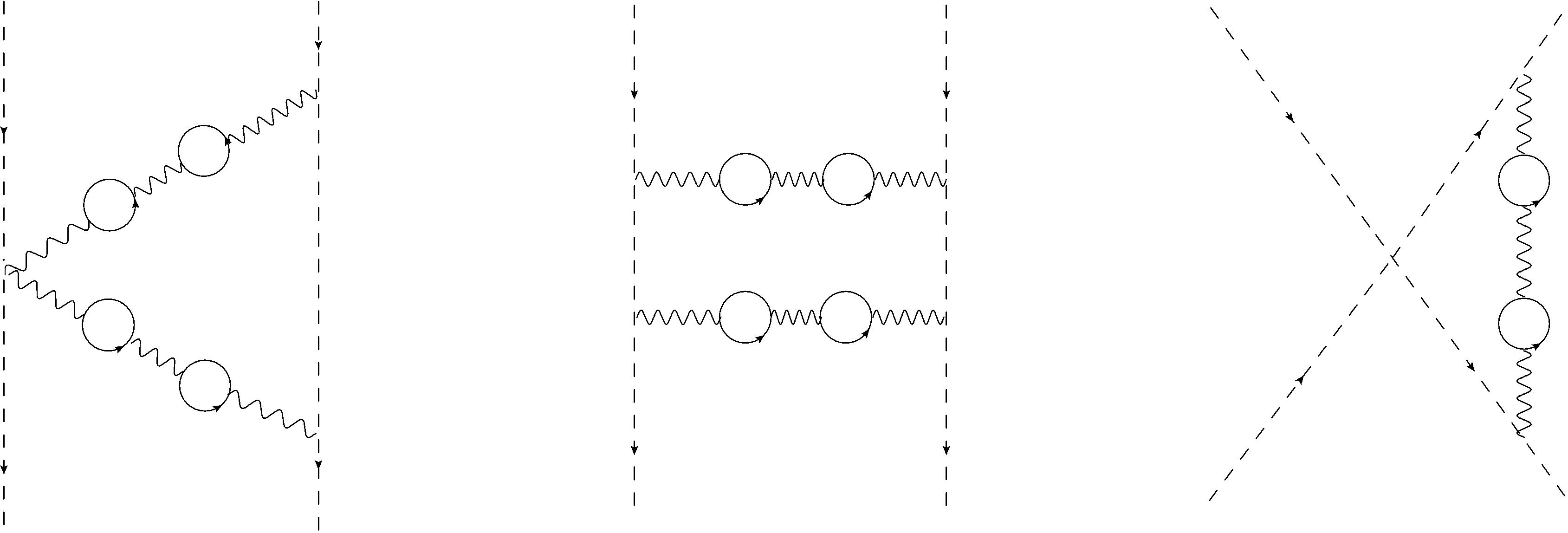}\hspace{0.05\columnwidth}
	\caption{$g^4$ with three point vertex on external scalar leg.}
	\label{fig:g_4_three_point}
\end{figure}

We  evaluate the leading order gauge vertex contribution to the scalar self coupling. Such contributions first appear at $ 1/{\mathcal{N}}^{\,2} $, and in the Landau gauge the only contribution stems from the diagram of Fig.~\ref{fig:g_4}. All other types of diagrams, see Fig.~\ref{fig:g_4_three_point}, contain at least one three-point gauge insertion on an external scalar leg. Since the gauge propagator is transverse in the Landau gauge, any such coupling will be proportional to the external momenta and vanish when this is taken to zero. Therefore these diagrams will not contribute to the vertex counter term. 

 We proceed by computing the diagrams in the Abelian theory before considering the non-Abelian one as well as  the semi-simple gauge groups. 

\subsubsection{Abelian computation}
In order to evaluate the vertex contribution due to the diagrams in Fig. \ref{fig:g_4}, we first denote by $ i\Lambda^{(n,m)} $ such a diagram with $ m  $ bubbles on the one propagator and $ n-m $ bubbles on the other. $ \Lambda^{(n,m)} $ and $ \Lambda^{(n,n-m)} $ are indistinguishable, therefore we include a factor of $ \tfrac{1}{2} $ for each pair $(n,m) $ to avoid double counting. This also agrees with the diagrams where $ n=2m $, in which case the two bubble chains are indistinguishable and they receive a symmetry factor $ \tfrac{1}{2} $ from the Feynman rules. In the limit of vanishing external momenta, all permutations of the scalar legs count the same.  Moreover, in this limit, only the loop momenta passes through the bubble chains and the loop integrals will be indifferent to which propagator the bubbles are placed on. The divergent part of the 4-point function at vanishing external momenta is obtained from   
	\begin{equation}\label{eq:quartic_full_sum}
	\pole{i \Lambda} = \sum_{n=0}^{\infty} \sum_{m=0}^{n} i \pole{\Lambda^{(n,m)}} \; + \; \mathrm{permutation} = 2\sum_{n=0}^{\infty} (n+1) \pole{i\Lambda^{(n,0)}}.
	\end{equation}
It is thus clear that it is sufficient to evaluate only the diagrams with bubbles on one of the gauge lines.
  
To regulate the IR-divergence of the relevant diagrams, we consider non-vanishing external momenta, as given in Fig. \ref{fig:g_4}, from which we obtain
	\begin{equation} \label{eq:quartic_diagram}
	i\Lambda^{(n,0)} = \dfrac{1}{2} \left(i 2 q_\phi^2 \tilde{g}_0^2\right)^2 \mu^{2\epsilon} \int\dfrac{\dd^d k}{(2\pi)^d} D^{\mu\nu}(p-k) D^{(n)}_{\nu\mu}(k) \ ,
	\end{equation} 
where the factor $ \tfrac{1}{2} $ is the aforementioned symmetry factor. Evaluating the integral one obtains the result
	\begin{equation} \label{eq:quartic_diagram_m=0}
	\begin{split}
	i\Lambda^{(n,0)} &= \dfrac{i\pi^2}{4 \mathcal{N}^2} \dfrac{q_\phi^4}{q_\Psi^4} (-2 K_0)^{n+2} \mu^{\epsilon} \left( -\dfrac{4\pi\mu^2}{p^2} \right)^{(n+1)\epsilon/2} \Gamma^n_0(\epsilon) \Gamma_\lambda(n,\epsilon) \ ,
	\end{split} 
	\end{equation}
with 
	\begin{equation}
	\Gamma_\lambda(n,\epsilon) = \dfrac{(3-\epsilon)(4 - \epsilon + n\epsilon)}{(n+1)\epsilon} \dfrac{\Gamma(1 - \tfrac{n+1}{2} \epsilon) \Gamma(1 -\tfrac{\epsilon}{2}) \Gamma(1 + \tfrac{n+1}{2}\epsilon) }{ \Gamma(2 + \tfrac{n}{2}\epsilon) \Gamma(2 - \tfrac{n+2}{2}\epsilon)} \ .
	\end{equation}
At this point we may sum over all the different diagrams as indicated by Eq. \eqref{eq:quartic_full_sum} to obtain the pole structure of the vertex.   By redefining the summation from $ n \longrightarrow n-2 $, one finds
	\begin{equation}
	i\pole{\Lambda} = \pole{\dfrac{i 54 \pi^2 \mu^\epsilon}{\mathcal{N}^2} \dfrac{q_\phi^4}{q_\Psi^4} \sum_{n=2}^{\infty} \left(-\dfrac{2K_0}{3} \right)^{n} \dfrac{1}{\epsilon^{n-1}} H_\lambda(n,\epsilon)} \ ,
	\end{equation} 
where
	\begin{equation}
	H_\lambda(n,\epsilon) =  \left(-\dfrac{4\pi \mu^2}{p^2} \right)^{(n-1)\epsilon/2}  \left[3\epsilon \, \Gamma_0(\epsilon) \right]^{n-2} \dfrac{(n-1)\epsilon}{12} \, \Gamma_\lambda(n-2, \epsilon) \ .
	\end{equation}
It is now possible to resum the pole structure of the vertex contribution using \eqref{eq:second_resum}. The resulting leading order gauge contribution to the quartic counter term pole is
	\begin{equation}
	\delta \lambda^{(1)} = \pole{\Lambda} = \dfrac{ 24 \pi^2 }{ \mathcal{N}^2} \dfrac{q_\phi^4}{q_\Psi^4}\,\mu^\epsilon\, K^2 H_\lambda(1,\tfrac{2}{3} K) \ ,
	\end{equation} 
with
	\begin{equation}
	H_\lambda(1,x) = \dfrac{(1 - \tfrac{x}{3}) \Gamma(4-x)}{ 6\Gamma^3(2 - \tfrac{x}{2}) \Gamma(1 + \tfrac{x}{2})} \ .
	\label{sum_quartic}
	\end{equation}

\subsubsection{Non-Abelian case} 
 In the non-Abelian case we have 
\begin{equation}
i\Lambda^{(n,0)}\udindices{ab}{cd} = \dfrac{1}{2} \left(i \tilde{g}_0^2 \right)^2 \braces{T_\phi^{A}, T_\phi^{B}}\udindices{a}{c} \braces{T_\phi^{A}, T_\phi^{B}}\udindices{b}{d} \,  \mu^{2\epsilon} \int\dfrac{\dd^d k}{(2\pi)^d} D^{\mu\nu}(p-k) D^{(n)}_{\nu\mu}(k) \ .
\end{equation} 
By  comparing this expression with the Abelian diagram of \eqref{eq:quartic_diagram} we can read off the non-Abelian result, paying attention to the fact that the color structure changes depending on the permutation of the external scalars. The contribution to the counter term in the non-Abelian theory is thus given by
	\begin{equation}
	\delta\lambda^{(1)} \udindices{ab}{cd} = \dfrac{ 24 \pi^2 d^2(R_\Psi) }{ \mathcal{N}^2 S_2^2(R_\Psi)} A\udindices{ab}{cd}\, K^2\, H_\lambda(1,\tfrac{2}{3} K) \ ,
	\end{equation}
	with 
\begin{equation}
	A\udindices{ab}{cd} = \dfrac{1}{8} \left(\braces{T_\phi^{A}, T_\phi^{B}}\udindices{a}{c} \braces{T_\phi^{A}, T_\phi^{B}}\udindices{b}{d} + \braces{T_\phi^{A}, T_\phi^{B}}\udindices{a}{d} \braces{T_\phi^{A}, T_\phi^{B}}\udindices{b}{c} \right) \ .
	\end{equation}

\subsubsection{Semi-simple gauge group}
\begin{figure}[t]
	\centering
	\includegraphics[width=0.6\columnwidth]{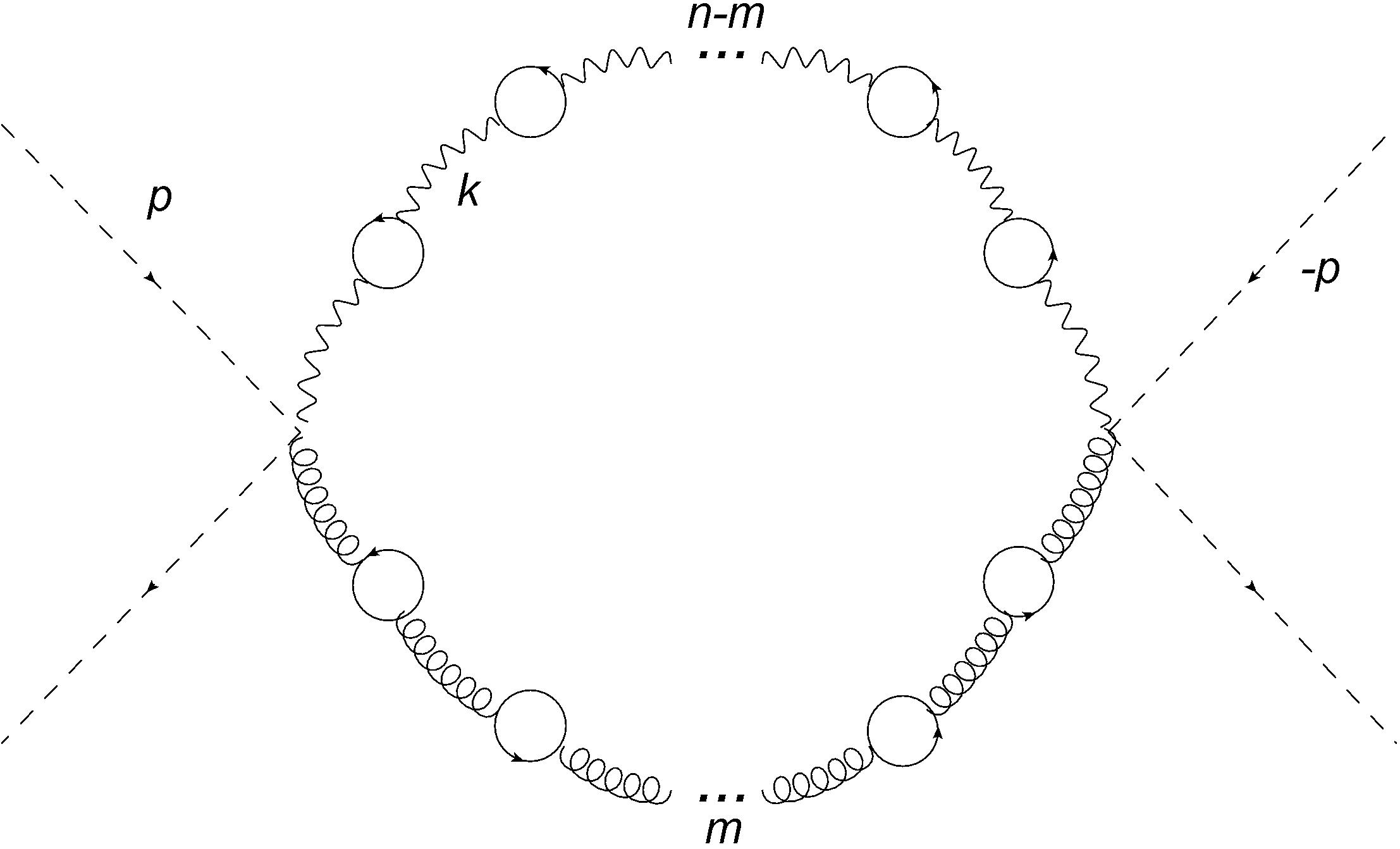}
	\caption{The mixed gauge term contributing to the quartic vertex in a semi-simple gauge theory.}
	\label{fig:g_4_semi-simple}
\end{figure}
The quartic coupling contains mixed gauge-coupling contributions already at LO in $ 1/\mathcal{N} $. If we consider the case where both the scalar and the vector-like fermions are charged under a semi-simple gauge group, then the quartic coupling receives mixed contributions of the type sketched in Fig. \ref{fig:g_4_semi-simple}. For every pair of simple gauge groups $ (G_\alpha, G_\beta) $, all the diagrams contain at least one power of  $ K_{\alpha,0} $ and $ K_{\beta,0} $ respectively. Starting from the simple diagram where all fermion bubbles are put on  the  $ G_\beta $ gauge line, we have 
	\begin{equation}
	i\Lambda^{(n,0)} =  (i 2 \tilde{g}_{\alpha,0} \tilde{g}_{\beta,0})^2 ( T^{A}_{\phi,\alpha} T^{B}_{\phi,\beta} )\udindices{a}{c} (T^{A}_{\phi,\alpha} T^{B}_{\phi,\beta} )\udindices{b}{d} \, \mu^{2\epsilon} \int\dfrac{\dd^d k}{(2\pi)^d} D_\alpha^{\mu\nu}(p-k) D^{(n)}_{\beta,\nu\mu}(k). 
	\end{equation}
Recall here that the generators belong to different gauge groups and therefore they commute. Comparing this diagram to Eqs. \eqref{eq:quartic_diagram} and \eqref{eq:quartic_diagram_m=0}  leads to 
	\begin{equation}
	\begin{split}
	i\Lambda^{(n,0)} = i ( T^{A}_{\phi,\alpha} T^{B}_{\phi,\beta} )\udindices{a}{c} & (T^{A}_{\phi,\alpha} T^{B}_{\phi,\beta})\udindices{b}{d} \dfrac{\pi^2 \mu^\epsilon}{2  \mathcal{N}^2} \dfrac{d(R_\Psi^{\alpha})\, d(R_\Psi^{\beta})}{S_2(R_\Psi^{\alpha}) \, S_2(R_\Psi^{\beta})} \\
	&(-2K_{\alpha,0}) (-2K_{\beta,0})^{n+1} \left(-\dfrac{4\pi\mu^2}{p^2} \right)^{(n+1)\epsilon/2} \Gamma_0^n(\epsilon) \Gamma_\lambda(n,\epsilon).
	\end{split}
	\end{equation}
We can distribute the fermion bubbles  in several ways on the two gauge lines. Each bubble gives a factor of $ K_{\alpha,0} $ or $ K_{\beta,0} $ depending on the gauge line, but the kinematic part of the diagram remains unchanged in the limit of vanishing external momentum. Because the pole structure does not depend on the external momentum, the pole structure of a general diagram can be related to the diagram with fermion bubbles on only one of the gauge line. In particular 
	\begin{equation}
	\pole{i\Lambda^{(n,m)}} = i \pole{\left(\dfrac{K_{\alpha,0}}{K_{\beta,0}} \right)^{m} \Lambda^{(n,0)}}. 
	\end{equation}
Thus summing over all possible bubbles and taking into account the different permutations of the external scalars, we arrive at 
	\begin{equation} \label{eq:semi-simple_quartic_sum}
	\begin{split}
	\pole{i \Lambda\udindices{ab}{cd}} = i& B_{\alpha,\beta}\udindices{ab}{cd} \dfrac{108 \pi^2}{\mathcal{N}^2} \dfrac{d(R_\Psi^{\alpha})\, d(R_\Psi^{\beta})}{S_2(R_\Psi^{\alpha}) \, S_2(R_\Psi^{\beta}) } \\
	&\qquad \times \pole{\sum_{n=1}^{\infty}\sum_{m=0}^{n-1} \left(-\dfrac{2K_{\alpha,0}}{3} \right)^{1+m} \left(-\dfrac{2 K_{\beta,0}}{3}\right)^{n-m} \dfrac{H_\lambda(n+1,\epsilon)   }{n\epsilon^n}},  
	\end{split}
	\end{equation}
having defined the tensor 
	\begin{equation}
	B_{\alpha,\beta}\udindices{ab}{cd} = \tfrac{1}{2} \left[( T^{A}_{\phi,\alpha} T^{B}_{\phi,\beta} )\udindices{a}{c} (T^{A}_{\phi,\alpha} T^{B}_{\phi,\beta} )\udindices{b}{d} + ( T^{A}_{\phi,\alpha} T^{B}_{\phi,\beta} )\udindices{a}{d} (T^{A}_{\phi,\alpha} T^{B}_{\phi,\beta} )\udindices{b}{c} \right].
	\end{equation} 
Employing Eq.~\eqref{eq:thrid_resum} and collecting the contributions from all the mixed terms, we find	
	\begin{equation}
	\begin{split}
	\delta \lambda^{(1)}\udindices{ab}{cd} =& \sum_{\alpha} \dfrac{ 24\pi^2 d^2(R_\Psi^{\alpha}) }{ \mathcal{N}^2\, S_2^2(R_\Psi^{\alpha})} A_1\udindices{ab}{cd}\, K_\alpha^2 H_\lambda(1,\tfrac{2}{3} K_\alpha)\\
	& +\sum_{\alpha < \beta} B_{\alpha, \beta}\udindices{ab}{cd} \dfrac{48 \pi^2}{\mathcal{N}^2} \dfrac{d(R_\Psi^{\alpha})\, d(R_\Psi^{\beta})}{S_2(R_\Psi^{\alpha}) \, S_2(R_\Psi^{\beta})} \dfrac{K_\alpha K_\beta}{K_\alpha -K_\beta} \int_{K_\beta}^{K_\alpha} \dd x \,H_\lambda(1, \tfrac{2}{3}x) \ ,
	\end{split}
	\end{equation}
which naturally also contains the unmixed contributions.

\subsection{Complete set of large $\mathcal{N}$ beta functions} 
Having evaluated all relevant diagrams we now compute all the beta functions using Eq.~\eqref{eq:beta_formulas}, starting with the Yukawa that reads   
\begin{equation}
\begin{split}
(\beta_y)_{aij} =& \frac{1}{32 \pi^2} \left[ (y_{b} y^{\dagger, b} y_{a})_{ij} + (y_{a} y^{\dagger, b} y_{b})_{ij} + 2\Tr[y_a y^{\dagger,b}] y_{bij} \right] \\
&- y_{aij} \sum_\alpha \frac{3\, d(R^\alpha_\Psi)}{16 \mathcal{N}} \frac{C_2(R_{\chi}^\alpha ) + C_2( R_{\xi}^\alpha)}{S_2(R_{\Psi}^\alpha ) } K_\alpha\, H_\psi (0,\tfrac{2}{3} K_\alpha) \\
&- y_{aij} \sum_\alpha \dfrac{3\, d(R_\Psi^\alpha)}{4\mathcal{N}} \frac{C_2(R_{\phi}^\alpha)}{S_2(R_{\Psi}^\alpha)} K_\alpha\, H_\phi(0, \tfrac{2}{3}K_\alpha) \\
& - y_{aij} \sum_\alpha \dfrac{3\, d(R_\Psi^\alpha)}{2\mathcal{N}} \frac{C_2(R_{\chi}^\alpha) + C_2(R_{\xi}^\alpha) - C_2(R_{\phi}^\alpha)}{2 S_2(R_{\Psi}^\alpha)} K_\alpha\, H_y(0,\tfrac{2}{3} K_\alpha) + \mathcal{O}\left(\frac{1}{\mathcal{N}^2}\right) \ . 
\end{split}
\label{total Yukawa}
\end{equation}
This result includes the 1-loop terms from the matter sector. 

\bigskip
For the quartic scalar coupling, the beta function is  
\begin{equation}
\begin{split}
\beta_\lambda\udindices{ab}{cd} =& \dfrac{1}{16\pi^2}\left(2 \lambda\udindices{ae}{cf} \lambda\udindices{bf}{de} +2 \lambda\udindices{ae}{df} \lambda\udindices{bf}{ce} + \lambda\udindices{ab}{ef} \lambda\udindices{ef}{cd} \right) +\frac{1}{4\pi^2} \Tr[y_d y^{\dagger,e}] \lambda\udindices{ab}{ce}\\
&- \frac{1}{4 \pi^2} \text{Tr}\left[y^a y^{\dagger}_c y^b y^{\dagger}_d + y^a y^{\dagger}_d y^b y^{\dagger}_c \right] - \lambda\udindices{ab}{cd} \sum_{\alpha} \dfrac{3\, d(R_\Psi^\alpha)}{ \mathcal{N} }\frac{C_2( R_{\phi}^\alpha)}{S_2( R_{\Psi}^\alpha)}  K_\alpha\, H_\phi(0, \tfrac{2}{3} K_\alpha)\\
&+ \dfrac{48 \pi^2}{\mathcal{N}^2} \sum_{\alpha< \beta} B_{\alpha,\beta}\udindices{ab}{cd}  \dfrac{d(R_\Psi^{\alpha})\, d(R_\Psi^{\beta})}{S_2(R_\Psi^{\alpha}) \, S_2(R_\Psi^{\beta})} \dfrac{K_\alpha K_\beta}{K_\alpha - K_\beta} \left[K_\alpha \, H_\lambda(1, \tfrac{2}{3} K_\alpha) - K_\beta \, H_\lambda(1, \tfrac{2}{3}K_\beta) \right] \\
&+ \dfrac{24 \pi^2}{\mathcal{N}^2} \sum_{\alpha} A_\alpha\udindices{ab}{cd} \dfrac{ d^2(R_\Psi^\alpha) }{ S_2^2(R_\Psi^\alpha)}\, \left[K_\alpha^2\, H_\lambda(1,\tfrac{2}{3} K_\alpha) + K_\alpha^3\,\dfrac{\partial}{\partial K_\alpha} H_\lambda(1,\tfrac{2}{3} K_\alpha)\right] + \mathcal{O}\left(\frac{1}{\mathcal{N}^3}\right) \ .
\end{split}
\label{total Quartic}
\end{equation}

\bigskip
Finally the gauge beta function for the full model  of  Eq.~\eqref{eq:L_model} is  
\begin{equation}\begin{split}
\beta_{K_\alpha} =  \dfrac{2K_\alpha^2}{3} & \left[ 1 +  \frac{d(R_\Psi^\alpha) }{\mathcal{N} S_2(R_{\Psi}^\alpha )} \left( \frac{1}{2} S_2( R_\chi^\alpha) + \frac{1}{2} S_2( R_\xi^\alpha) + \frac{1}{4} S_2(R_{\phi}^\alpha) \right) \right. \\
&\qquad + \frac{d(G)}{\mathcal{N}} H_1^{(\alpha)} (K_\alpha) + \dfrac{1}{\mathcal{N}} \sum_{\beta\neq \alpha} d(G_\beta) F_1(K_\beta) \bigg] + \mathcal{O}\left(\frac{1}{\mathcal{N}^2}\right) \ .
\label{total Gauge}
\end{split}\end{equation}
\noindent
We note that the $1/{\mathcal{N}}$ counting is consistent with the fact that $\lambda \sim y^2 \sim g^2 \sim 1/{\mathcal{N}}$.

\subsection{A mnemonic for Yukawa and quartic beta functions } 
Since the beta functions of many phenomenological models are known to LO, it is convenient to rewrite the above Yukawa and quartic beta function Eq.~\eqref{total Yukawa} and Eq.~\eqref{total Quartic} in a more compact form. With this prescription one can immediately obtain the bubble diagram contributions to known 1-loop beta functions by simply using the following recipe. 
The Yukawa beta function at large number of fermions can be written in the following compact form
\begin{equation}
\beta_y=c_1y^3 + y \sum_{\alpha}c_\alpha K_\alpha I_y\left(K_\alpha\right),\,
\label{eq-simplifiedyukawa}
\end{equation}
with 
\begin{equation}
I_y\left(K_\alpha\right)=H_\phi\left(0,\tfrac{2}{3}K_\alpha\right)\left(1+K_\alpha\frac{C_2\left(R_\phi^\alpha\right)}{6\left(C_2\left(R_{\chi}^\alpha\right)+C_2\left(R_{\xi}^\alpha\right)\right)}\right)\,,
\end{equation}
containing information about the resummed fermion bubbles and $c_1,\,c_\alpha$ are the standard 1-loop coefficients for the Yukawa beta function. Thus, when $c_1,\,c_\alpha$ are known, the total Yukawa beta function with bubble diagram contributions is straightforward.
Similarly, for the quartic coupling we write
\begin{equation}
\beta_\lambda=c_1\lambda^2+\lambda \sum_{\alpha}c_\alpha K_\alpha I_{\lambda g^2}\left(K_\alpha\right)+\sum_{\alpha} c'_\alpha K_\alpha^2I_{g^4} \left(K_\alpha\right) +\sum_{\alpha < \beta}c_{\alpha\beta} K_\alpha K_\beta I_{g_1^2g_2^2}\left(K_\alpha,\,K_\beta\right)\,,
\end{equation}
with $c_1,\,c_\alpha,\,c'_\alpha,\,c_{\alpha\beta}$  the known 1-loop coefficients\footnote{Clearly these are not the same numerical coefficients appearing in \eqref{eq-simplifiedyukawa}.} for the quartic beta function and the resummed fermion bubbles appear via 
\begin{equation}
\begin{split}
I_{\lambda g^2}\left(K_\alpha\right) &=H_\phi\left(0,\tfrac{2}{3}K_\alpha\right)\\
I_{g^4}\left(K_\alpha\right)&=H_\lambda\left(1,\tfrac{2}{3}K_\alpha\right)+K_\alpha\frac{dH_\lambda\left(1,\tfrac{2}{3}K_\alpha\right)}{dK_\alpha}\\
I_{g_1^2g_2^2}\left(K_\alpha,\,K_\beta\right)&=\frac{1}{K_\alpha-K_\beta}\left[K_\alpha H_\lambda\left(1,\tfrac{2}{3}K_\alpha\right)-K_\beta H_\lambda\left(1,\tfrac{2}{3}K_\beta\right)\right]\,.
\end{split}
\end{equation} 
It is thus also straightforward to obtain the total quartic beta function including the bubble diagram contributions when $c_1,\,c_\alpha,\,c_\alpha',\,c_{\alpha\beta}$ are known.

\subsection{Pole structure of the beta functions}
We now elucidate the pole structure of the resummed beta functions which is a characteristic feature of the theories investigated here.  

Since the pole structure of beta function in theories with a simple gauge group has been discussed already in literature~\cite{Holdom:2010qs,Pica:2010xq,Shrock:2013cca,Antipin:2017ebo}, we move immediately to consider the semi-simple gauge-fermion theories. Here we observe that if the group structures contains an Abelian factor  the corresponding beta function is such that it still features a singularity for  $K=\frac{15}{2}$ regardless of the presence of other non-Abelian factors. This is so since the extra contribution assume the form of $F_1$ rather than $H_1$ (see Eq.~\eqref{total Quartic}). Thus, it is not possible to shift the resulting UV fixed point value of the Abelian gauge coupling away from the Abelian pole. This is clearly manifest in the phase diagram structure of Fig.~\ref{Phase1}.

\bigskip 

For the Yukawa beta function we first observe, using Eqs.~\eqref{sum_fermion},  \eqref{sum_scalar}, and \eqref{sum_yukawa},  that the following relations hold:
\begin{equation}
	H_{\psi}(0,x) = x H_0(x), \quad H_{\phi}(0, x) = H_0(x), \quad H_y(0,x) = H_\lambda(1,x) = (1-\tfrac{x}{4}) H_0(x),
	\end{equation}
	where 
	\begin{equation}\label{key}
	H_0(x) = \dfrac{(1 - \tfrac{x}{3}) \Gamma(4-x)}{3 \Gamma^2(2 - \tfrac{x}{2}) \Gamma(3 - \tfrac{x}{2}) \Gamma(1 + \tfrac{x}{2})} \ .
	\end{equation}
This means that they all inherit a pole at $x=5$ yielding 
\begin{equation}
H_\psi\left(0,\tfrac{2}{3}K_\alpha\right)\sim\frac{1}{\frac{15}{2}-K_\alpha},\quad H_\phi\left(0,\tfrac{2}{3}K_\alpha\right)\sim\frac{1}{\frac{15}{2}-K_\alpha},\quad H_y\left(0,\tfrac{2}{3}K_\alpha\right)\sim\frac{1}{K_\alpha-\frac{15}{2}}\,.
\end{equation}
Thus the Yukawa coupling RG function Eq.~\eqref{total Yukawa} near the pole will assume the following form:
\begin{equation}
\beta_y=c_1y^3 + y K_\alpha\left(\frac{1}{K_\alpha-\frac{15}{2}}\right)\left(c_2+c_3K_\alpha\right)\,,\label{yukawa_RG_pole}
\end{equation}
where $c_1,\,c_2,\,c_3$ are positive constants stemming from the group structure of the theory. 
It is clear that the three summation functions altogether provide large negative contributions when approaching the pole (i.e.~$K_\alpha=\frac{15}{2}$) from the left. The pole in the Yukawa beta function appears at the original Abelian gauge coupling location (see Eq.~\eqref{abelian_summation}). This implies that if the gauge group features an Abelian factor, from  Eq.~\eqref{yukawa_RG_pole},  we deduce that the Yukawa gauge coupling vanishes in the UV (free rather than safe). The situation changes dramatically when only non-Abelian gauge groups are involved. This is so since the non-Abelian gauge beta function reaches an UV fixed point  at $K_\alpha=3$, which is clearly away from the Abelian pole, allowing for non-trivial UV zeros of the Yukawa beta function.  

\bigskip
Similarly to the Yukawa beta function the RG equation for the quartic coupling, due to the $K_\alpha H_\phi\left(0,\frac{2}{3}K_\alpha\right)$ term in Eq.~\eqref{total Quartic}, receives a large negative contribution at the Abelian pole. 
This is made explicit by the relations  
\begin{equation}
H_\lambda\left(1,\tfrac{2}{3}K_\alpha\right)=H_y\left(0,\tfrac{2}{3}K_\alpha\right)\sim\frac{1}{K_\alpha-\frac{15}{2}},\qquad\frac{\partial}{\partial K_\alpha}H_\lambda\left(1,\tfrac{2}{3}K_\alpha\right)\sim-\frac{1}{\left(K_\alpha-\frac{15}{2}\right)^2}\,.
\end{equation}
Thus, the  quartic coupling RG function Eq.~\eqref{total Quartic} near the first singularity assumes the form
\begin{equation}
\beta_\lambda=c_1\lambda^2+c_2\lambda K_\alpha \left(\frac{1}{K_\alpha-\frac{15}{2}}\right)+c_3 K_\alpha^2\left(\frac{1}{K_\alpha-\frac{15}{2}}-\frac{1}{\left(K_\alpha-\frac{15}{2}\right)^2}\right)\,,\label{quartic_RG_pole}
\end{equation}
where $c_1,\,c_2,\,c_3$ denote positive constants (distinct from the Yukawa case). From Eq.~\eqref{quartic_RG_pole} we learn that the quartic beta function has an even more singular structure located at the Abelian pole. If the theory contains an Abelian gauge group one observes that the quartic coupling develops an explosive behavior ($\lambda \propto \exp(N_f)$) at the Abelian fixed point and the fixed point analysis cannot be trusted. 

The situation for the non-Abelian case resemble the Yukawa case. Here the non-Abelian UV fixed point is achieved at $K_\alpha=3$ which is below and sufficiently away from the pole in the quartic coupling, allowing for (depending on the theory) the existence of UV fixed points in all couplings.

\section{Conclusion} \label{sec:conclusion}
We investigated gauge-Yukawa theories  at large number of  gauged fermion fields. We begun our analysis by reviewing the state-of-the-art of the  gauge-fermion theories. We considered also semi-simple groups and by discussing their RG phase diagram we discovered a complete asymptotically safe fixed point  which turns out to be repulsive in all gauge couplings.

Subsequently  we enriched the original gauge-fermion theories by  introducing two  Weyl gauged fermions  transforming according to arbitrary representations of the gauge group and further added a complex gauged scalar. The latter is responsible for the presence of Yukawa and quartic scalar self-coupling interactions. On par with the gauge sector, we  determined  the leading $ 1/N_f $ Yukawa and quartic beta functions. We then discussed the pole structure of the system of RG equations. This has an immediate impact on the existence, location and stability of  related fixed points.  In particular one observes that when an Abelian gauge coupling is present in the theory, the Yukawa beta function is driven to be free while the quartic coupling becomes uncontrollable, de facto requiring a fully non-perturbative analysis near this point. The situation changes dramatically when only non-Abelian gauge couplings are present. Because the latter achieve a fixed point at a much lower value of the Abelian one (still appearing as the only pole in the Yukawa and quartic beta function) Yukawa and quartic couplings (depending on the theory) can still admit UV interacting fixed points. These results cannot be extended to the supersymmetric case \cite{Intriligator:2015xxa} for a number of reasons.  The first reason is that the resummation procedure would have to respect supersymmetry and, in addition, it has already been proven in \cite{Intriligator:2015xxa} that it is impossible to have an UV fixed point for any $N_f$ in super QCD. 

Our work elucidates, corrects, consolidates, and extends results obtained earlier in the literature \cite{PalanquesMestre:1983zy,Gracey:1996he,Holdom:2010qs,Pica:2010xq,Shrock:2013cca,Mann:2017wzh,Pelaggi:2017abg,Antipin:2017ebo,Kowalska:2017pkt}. It also provides the stepping stone and the needed instruments for future theoretical and phenomenological extensions and analyses.

\subsection*{Acknowledgments}
The work has been partially supported by the Danish National Research Foundation under grant {DNRF:90, the Croatian Science Foundation under the project 4418 and the Natural Sciences and Engineering Research Council of Canada (NSERC). OA also acknowledges the partial support by the H2020 CSA Twinning project No.692194, RBI- T-WINNING.  Z.W.~Wang thanks Robert Mann, Tom Steele, Cacciapaglia Giacomo and Emiliano Molinaro for helpful suggestions.

\appendix
\section{Resummation formulas} \label{sec:appendix}
Here we present proofs for the four resummation formulas used for the large $ \mathcal{N} $ computations. Regardless of the quantity in question only the pole structure at $ \epsilon \rightarrow 0 $ will be relevant. We will use the notation ``$ \underset{\epsilon\rightarrow 0}{\sim} $'' to mean equal divergent parts. Note that all of these resummations are only valid to LO in $ 1/\mathcal{N} $.

All the resummation formulas rely on a function $ H(n,\epsilon) $ being regular both for $ \alpha = n\epsilon \rightarrow 0 $ with $ \epsilon $ constant and for $ \epsilon\rightarrow 0 $ with $ \alpha $ constant. In general, the functions occurring in this paper can easily be checked to satisfy this condition, except, possibly, for the term $ [3\epsilon \, \Gamma_0(\epsilon)]^{n} $ with fixed $ \alpha $ and $ \epsilon \rightarrow 0 $. To see that this term is well behaved, it is sufficient to note that the base satisfy  
	\begin{equation}
	\left(\dfrac{6 \Gamma^2(2-\tfrac{\epsilon}{2}) \Gamma(1 + \tfrac{\epsilon}{2}) }{\Gamma(4-\epsilon)} \right)^{\alpha/\epsilon - 1} = \left(1 + \epsilon \,f(\epsilon) \right)^{\alpha/\epsilon}, 
	\end{equation}
for some regular function $ f $ that is regular in $ 0 $. To prove that this expression has no pole, observe that there must exist constants $ c, \delta >0 $ such that 
	\begin{equation}
	\abs{f(\epsilon)} \leq c, \qquad \forall \abs{\epsilon} < \delta.  
	\end{equation} 
It must then hold that 
	\begin{equation}
	\lim_{\epsilon\rightarrow 0} \abs{1 + \epsilon\, f(\epsilon)}^{\alpha/\epsilon} \leq \lim_{\epsilon\rightarrow 0_+} \left(1 +c\, \epsilon\right)^{\alpha/\epsilon} = \lim_{x \rightarrow \infty} \left(1 + \dfrac{c}{x} \right)^{x \alpha} = e^{c \alpha}
	\end{equation}
having substituted $ x= \tfrac{1}{\epsilon} $. It is thus found that $ [3\epsilon \, \Gamma_0(\epsilon)]^{n} $ is without a pole in $ \epsilon \rightarrow 0 $ and fixed $ n\epsilon $.

\subsection{First resummation}
	\begin{equation}\label{eq:fourth_resum}
	\boxed{\sum_{n=1}^{\infty} \left( -\frac{2K_0}{3} \right)^n \frac{H(n+1,\epsilon)}{(n+1) \epsilon^{n}}  \underset{1/\epsilon}{\sim} -\dfrac{2}{3\epsilon} \int_0^{K} \dd x\, \left(1 - \dfrac{x}{K}\right) H(0, \tfrac{2}{3}x) }
	\end{equation}
The proof, originally due to \cite{PalanquesMestre:1983zy}, is presented here to keep the work self-contained. We define 
	\begin{equation} \label{eq:first_resum_start}
	R(\epsilon) = \sum_{n=1}^{\infty} \left( -\frac{2K_0}{3} \right)^n \frac{H(n+1,\epsilon)}{(n+1) \epsilon^{n}} .
	\end{equation}
The resummation proceeds assuming that $ H(n,\epsilon) $ can be written as a power series
	\begin{equation} \label{eq:nepsilon_expansion}
	H(n,\epsilon) = \sum_{j=0}^{\infty} H^{(j)}(\epsilon) (n\epsilon)^{j}, 
	\end{equation}  
where every $ H^{(j)}(\epsilon) $ are regular in $ \epsilon $. Such an expansion only exists, if $ H(\tfrac{\alpha}{\epsilon}, \epsilon) $ is regular in both $ \alpha $ and $ \epsilon $. Furthermore, the bare coupling is renormalized with $ K_0 = Z_K^{-1} K $. As this is a LO in $ 1/\mathcal{N} $ computation, one may then expand 
	\begin{equation} \label{eq:ZK_n}
	Z_K^{-n} = \left[1 - \dfrac{2K}{3\epsilon} + O\left(\dfrac{1}{\mathcal{N}} \right) \right]^{-n} = \sum_{k = 0}^{\infty} \binom{n + k -1}{k} \left(\dfrac{2K}{3\epsilon}\right)^k + O\left(\dfrac{1}{\mathcal{N}}\right). 
	\end{equation}
Inserting all this back into Eq. \eqref{eq:first_resum_start}, we find
	\begin{equation}
	R(\epsilon) = \sum_{n=1}^{\infty} \sum_{k=0}^{\infty} \sum_{j=0}^{\infty} \left( - \frac{2K}{3} \right)^{n+k} \binom{n + k - 1}{k}  \frac{(-1)^k (n+1)^{j-1} }{\epsilon^{n+k-j}}  H^{(j)}(\epsilon), 
	\end{equation}
where $ H^{(j)}(\epsilon) $ are regular functions of $ \epsilon $. Defining $ m = n + k $, the sums are redefined in terms of $ m $ and $ k $. The only poles in $ \epsilon $ occur for $ j < m  $, so the divergent part of $ R $ is given by
	\begin{equation}
	R(\epsilon) \underset{\epsilon \rightarrow 0}{\sim} \sum_{m=1}^{\infty} \left( - \frac{2K}{3} \right)^{m} \sum_{j=0}^{m-1}  \dfrac{H_j(\epsilon)}{\epsilon^{m-j}} \sum_{k=0}^{m-1}  \binom{m - 1}{k}  (-1)^k (m - k + 1)^{j-1}.
	\end{equation}
The sum greatly simplifies as the identity 
	\begin{equation} \label{eq:binom_identity}
	\sum_{k = 0}^{m}  \binom{m}{k} (-1)^k (x -k)^{j} = 0 , 
	\end{equation}
valid for all integer $ 0\leq j < m $ and real numbers $ x $, implies that the $ k $ sum vanish for all $ j\neq 0 $. Meanwhile the $ j=0 $ term evaluates to 
	\begin{equation}
	\begin{split}
	& \sum_{k=0}^{m-1}  \binom{m - 1}{k} \dfrac{(-1)^k}{m-k+1} = \dfrac{(-1)^{m-1}}{m(m+1)}.	
	\end{split}
	\end{equation}
Performing the $ j $ and $ k $ sums then yield
	\begin{equation}
	R(\epsilon) \underset{\epsilon \rightarrow 0}{\sim} -\sum_{m = 1}^{\infty} \left( \frac{2K}{3} \right)^{m} \dfrac{H^{(0)}(\epsilon)}{\epsilon^{m}} \dfrac{1}{m(m+1)}.
	\end{equation}
Expanding now $ H^{(0)}(\epsilon) = \sum_{\ell=0}^{\infty} H_\ell^{(0)} \epsilon^\ell $ and selecting the simple pole one finds 
	\begin{equation}
	R(\epsilon) \underset{1/\epsilon}{\sim} - \dfrac{1}{\epsilon} \sum_{\ell = 0}^{\infty} \left( \frac{2K}{3} \right)^{\ell + 1} \dfrac{H^{(0)}_\ell }{(\ell+1)(\ell+2)}. 
	\end{equation}
Finally, before resumming the power series in $ H_\ell^{(0)} $, the fraction is decomposed so that
\begin{equation}
\begin{split}
R(\epsilon) & \underset{1/\epsilon}{\sim} -\dfrac{1}{\epsilon} \left[\sum_{\ell = 0}^{\infty} \left( \frac{2K}{3} \right)^{\ell + 1} \dfrac{H^{(0)}_\ell }{(\ell+1)} - \dfrac{3}{2K} \sum_{\ell = 0}^{\infty} \left( \frac{2K}{3} \right)^{\ell + 2} \dfrac{H^{(0)}_\ell }{(\ell+2)}\right] \\
& \underset{1/\epsilon}{\sim} -\dfrac{2}{3\epsilon} \left[\int_0^{K} \dd x \, H(0, \tfrac{2}{3}x) - \dfrac{1}{K} \int_0^{K} \dd x\, x H(0, \tfrac{2}{3}x) \right] \ , 
\end{split}
\end{equation}
taking a $ K $ derivative and resumming the power series before integrating again. This concludes the proof.

\subsection{Second resummation} 
	\begin{equation}\label{eq:first_resum}
	\boxed{ \sum_{n=1}^{\infty} \left(-\dfrac{2 K_0}{3}\right)^n \dfrac{1}{n\epsilon^n} H(n,\epsilon) \underset{1/\epsilon}{\sim} -\dfrac{2}{3\epsilon} \int_{0}^{K}\dd x\, H(0,\tfrac{2}{3}x)}
	\end{equation}	
To prove this formula~\cite{Kowalska:2017pkt}, let 
	\begin{equation}
	S(\epsilon) = \sum_{n=1}^{\infty} \left(-\dfrac{2 K_0}{3}\right)^n \dfrac{1}{n\epsilon^n} H(n,\epsilon) .
	\end{equation}
Assuming again that $ H(n, \epsilon) $ is sufficiently regular, it may be expanded according to Eq. \eqref{eq:nepsilon_expansion}. At the same time, the bare coupling is renormalized by $ K_0= Z_K^{-1} K $. Then using the expansion Eq. \eqref{eq:ZK_n} one finds that
	\begin{equation}
	S(\epsilon)  = \sum_{n=1}^{\infty} \sum_{k = 0}^{\infty} \sum_{j=0}^{\infty} \left(-\dfrac{2 K}{3}\right)^{n+k} \binom{n + k -1}{k} \dfrac{(-1)^k n^{j-1}}{ \epsilon^{n + k - j} } H^{(j)}(\epsilon).
	\end{equation}	
Note that as $ H^{j}(\epsilon) $ is regular in $ 0 $, only the terms in the sum with $ j\leq n + k - 1 $ will contribute to the pole structure. The sums are redefined with $ m = n + k $ and so
	\begin{equation}
	S(\epsilon) \underset{\epsilon \rightarrow 0}{\sim}  \sum_{m=1}^{\infty} \left(-\dfrac{2 K}{3}\right)^{m} \sum_{j=0}^{m-1} \dfrac{H^{(j)}(\epsilon)}{\epsilon^{m-j}} \sum_{k = 0}^{m-1}  \binom{m -1}{k} (-1)^k (m-k)^{j-1}.
	\end{equation}	
According to the identity \eqref{eq:binom_identity}, the $ k $ sum vanish for all $ j\neq 0 $. The sum is thus evaluated by keeping just the $ j=0 $ term\footnote{For $ j=0 $ one has to use the summation: \[\sum_{k=0}^{n} \binom{n}{k} \dfrac{(-1)^k}{n-k+1} = (-1)^n \sum_{k=0}^{n} \binom{n}{k} \dfrac{(-1)^k}{k+1} = \dfrac{(-1)^n}{n+1} \sum_{k=0}^{n} \binom{n+1}{k+1} (-1)^k = \dfrac{(-1)^n}{n+1} \] }
	\begin{equation}
	S(\epsilon) \underset{\epsilon \rightarrow 0}{\sim}  -\sum_{m=1}^{\infty} \left(\dfrac{2 K}{3}\right)^{m} \dfrac{1}{m} \dfrac{H^{(0)}(\epsilon)}{\epsilon^{m}} .
	\end{equation}
Expanding $ H^{(0)}(\epsilon) $ as power series, the sum gives 
	\begin{equation}
	S(\epsilon) \underset{\epsilon \rightarrow 0}{\sim}  -\sum_{m=1}^{\infty} \sum_{\ell=0}^{\infty} \left(\dfrac{2 K}{3}\right)^{m} \dfrac{1}{m} \dfrac{H^{(0)}_{\ell} }{\epsilon^{m-\ell}}, \qquad \mathrm{where}\quad  H^{(0)}(\epsilon)  = \sum_{\ell=0}^{\infty} H^{(0)}_{\ell} \epsilon^\ell .
	\end{equation}
As only the simple pole in epsilon is of interest to us, the resummation can now be concluded
	\begin{equation}
	S(\epsilon) \underset{1/\epsilon}{\sim} -\dfrac{1}{\epsilon} \sum_{\ell=0}^{\infty} \left(\dfrac{2 K}{3}\right)^{\ell +1} \dfrac{1}{\ell + 1} H^{(0)}_{\ell} \underset{1/\epsilon}{\sim} -\dfrac{2}{3\epsilon} \int_{0}^{K}\dd x \sum_{\ell=0}^{\infty} \left(\dfrac{2 x}{3}\right)^{\ell} H^{(0)}_{\ell} 
	\end{equation}
The second to last step is done by taking a $ K $ derivative before reintegrating so that the power series of $ H^{(0)}(x) = H(0, x)  $ can finally be resummed proving \eqref{eq:first_resum}.

\subsection{Third resummation} 
\begin{equation}\label{eq:second_resum}
\boxed{\sum_{n=2}^{\infty} \left(-\dfrac{2K_0}{3} \right)^{n} \dfrac{1}{\epsilon^{n-1}} H(n,\epsilon) \underset{1/\epsilon}{\sim} \dfrac{1}{\epsilon} \left(\dfrac{2K}{3} \right)^2 H(1,\tfrac{2}{3} K)}
\end{equation}
To prove that this is the case, let
\begin{equation}
T(\epsilon) = \sum_{n=2}^{\infty} \left(-\dfrac{2K_0}{3} \right)^{n} \dfrac{1}{\epsilon^{n-1}} H(n,\epsilon).
\end{equation} 
Here too, the function $ H(n,\epsilon) $ is expanded as a power series in $ n\epsilon $ as given in Eq.~\eqref{eq:nepsilon_expansion} such that all $ H^{(j)}(\epsilon) $ are regular in $ \epsilon = 0 $. Simultaneously the bare couplings are renormalized $ K_0 = Z_K^{-1} K_0 $ using Eq.~\eqref{eq:ZK_n}, and we write
\begin{equation}
T(\epsilon) = \sum_{n=2}^{\infty} \sum_{j=0}^{\infty} \sum_{k=0}^{\infty} \left(-\dfrac{2K}{3} \right)^{n+k} \binom{n+k-1}{k} (-1)^{k} H^{(j)}(\epsilon) \dfrac{n^{j}}{\epsilon^{n+k-j-1}}. 
\end{equation}
The sums can be rewritten in terms of $ m =n + k $ so that 
\begin{equation}
T(\epsilon) = \sum_{m=2}^{\infty} \left(-\dfrac{2K}{3} \right)^{m} \sum_{j=0}^{\infty}  \dfrac{H^{(j)}(\epsilon)}{\epsilon^{m-j-1}} \left[\sum_{k=0}^{m-1} \binom{m-1}{k} (-1)^{k} (m-k)^{j} - (-1)^{m-1}\right] . 
\end{equation}
Here the last terms is a compensation for the fact that the $ k $ sum is taken to go all the  way to $ m-1 $ rather than just to $ m-2 $. Poles in $ \epsilon $ will only appear for $ j\leq m-2 $ in which case the identity~\eqref{eq:binom_identity} causes the first term in the sum to vanish. Thus, for the pole structure it holds that
\begin{equation}
T(\epsilon) \underset{\epsilon\rightarrow0}{\sim}  \sum_{m=2}^{\infty} \left(\dfrac{2K}{3} \right)^{m} \sum_{j=0}^{\infty}  \dfrac{H^{(j)}(\epsilon)}{\epsilon^{m-j-1}} = \sum_{m=2}^{\infty} \left(\dfrac{2K}{3} \right)^{m} \dfrac{H(1, \epsilon)}{\epsilon^{m-1}},
\end{equation} 
where the last equality is due to a resummation of the $ H $ function. Re-expanding now $ H(1,\epsilon) $ yields
\begin{equation}
T(\epsilon) \underset{\epsilon\rightarrow0}{\sim}  \left(\dfrac{2K}{3} \right)^2 \sum_{m=0}^{\infty} \sum_{\ell=0}^{\infty} \left(\dfrac{2K}{3} \right)^{m} \dfrac{H_{\ell}}{\epsilon^{m+1-\ell}}, \qquad \mathrm{where} \quad H(1, \epsilon) = \sum_{\ell=0}^{\infty} H_{\ell} \epsilon^\ell.
\end{equation}
At this point one can immediately determine the simple $ \epsilon $ pole, which is evaluated to
\begin{equation}
T(\epsilon) \underset{1/\epsilon}{\sim} \dfrac{1}{\epsilon} \left(\dfrac{2K}{3} \right)^2 \sum_{m=0}^{\infty} \left(\dfrac{2K}{3} \right)^{m} H_{m} = \dfrac{1}{\epsilon} \left(\dfrac{2K}{3} \right)^2 H(1, \tfrac{2}{3} K). 
\end{equation} 

\subsection{Fourth resummation}
\begin{equation}\label{eq:thrid_resum}
\boxed{\sum_{n=1}^{\infty}\sum_{m=0}^{n-1} \left(-\dfrac{2K_{\alpha,0}}{3}\right)^{1+m} \left(-\dfrac{2K_{\beta,0}}{3}\right)^{n-m} \dfrac{H(n,\epsilon) }{n\epsilon^n}\underset{1/\epsilon}{\sim} \dfrac{4}{9\epsilon} \dfrac{K_\alpha K_\beta}{K_\alpha - K_\beta} \int_{K_\beta}^{K_\alpha} \dd x \,H(0,\tfrac{2}{3}x)}
\end{equation}
To prove this, denote by
	\begin{equation}
	U(\epsilon) = \sum_{n=1}^{\infty}\sum_{m=0}^{n-1} \left(-\dfrac{2 K_{\alpha,0}}{3}\right)^{1+m} \left(-\dfrac{2 K_{\beta,0}}{3}\right)^{n-m} \dfrac{H(n,\epsilon) }{n\epsilon^n}. 
	\end{equation}
First the sum over $ m $ is performed by noting that 
	\begin{equation}
	\begin{split}
	\sum_{m=0}^{n-1} &\left(-\dfrac{2K_{\alpha,0}}{3}\right)^{1+m} \left( -\dfrac{2 K_{\beta,0} }{3} \right)^{n-m} \\
	&\qquad = \dfrac{2}{3} \left[\dfrac{1}{K_{\alpha,0} } - \dfrac{1}{K_{\beta,0} }\right]^{-1} \left[\left( -\frac{2 K_{\alpha,0} }{3} \right)^n - \left(- \frac{2 K_{\beta,0}}{3} \right)^n \right].
	\end{split}
	\end{equation}
Now going from the bare to the renormalized couplings using $ K_{\alpha,0} = Z_{K_\alpha}^{-1} K_\alpha = K_\alpha (1 - \tfrac{2 K_\alpha}{3\epsilon})^{-1} $ at LO in $ 1/\mathcal{N} $, it is beneficial to first consider the first term,
	\begin{equation}
	\begin{split}
	&\left[\dfrac{1}{K_{\alpha,0} } - \dfrac{1}{K_{\beta,0} }\right]^{-1}  = - \dfrac{K_\alpha K_\beta}{K_\alpha - K_\beta}.
	\end{split}
	\end{equation}    
This term is thus finite and does not contribute to the pole structure. The sum can be written as
	\begin{equation}
	U(\epsilon) = -\dfrac{2}{3} \dfrac{K_\alpha K_\beta}{K_\alpha -K_\beta} \sum_{n=1}^{\infty} \left[ \left(-\frac{2K_{\alpha ,0}}{3}\right)^n - \left(-\frac{2K_{\beta, 0}}{3}\right)^n \right] \dfrac{H(n,\epsilon) }{n\epsilon^n}.
	\end{equation}
At this stage the resummation \eqref{eq:first_resum} can be applied directly to obtain 
	\begin{equation}
	U(\epsilon) \underset{1/\epsilon}{\sim} \dfrac{4}{9\epsilon} \dfrac{K_\alpha K_\beta}{K_\alpha - K_\beta} \left[\int_{0}^{K_\alpha} \dd x\, H(0,\tfrac{2}{3}x) - \int_{0}^{K_\beta} \dd x\, H(0, \tfrac{2}{3} x) \right].
	\end{equation}

\bibliographystyle{apsrev4-1}
\bibliography{references}

\begin{thebibliography}{39}%
\makeatletter
\providecommand \@ifxundefined [1]{%
 \@ifx{#1\undefined}
}%
\providecommand \@ifnum [1]{%
 \ifnum #1\expandafter \@firstoftwo
 \else \expandafter \@secondoftwo
 \fi
}%
\providecommand \@ifx [1]{%
 \ifx #1\expandafter \@firstoftwo
 \else \expandafter \@secondoftwo
 \fi
}%
\providecommand \natexlab [1]{#1}%
\providecommand \enquote  [1]{``#1''}%
\providecommand \bibnamefont  [1]{#1}%
\providecommand \bibfnamefont [1]{#1}%
\providecommand \citenamefont [1]{#1}%
\providecommand \href@noop [0]{\@secondoftwo}%
\providecommand \href [0]{\begingroup \@sanitize@url \@href}%
\providecommand \@href[1]{\@@startlink{#1}\@@href}%
\providecommand \@@href[1]{\endgroup#1\@@endlink}%
\providecommand \@sanitize@url [0]{\catcode `\\12\catcode `\$12\catcode
  `\&12\catcode `\#12\catcode `\^12\catcode `\_12\catcode `\%12\relax}%
\providecommand \@@startlink[1]{}%
\providecommand \@@endlink[0]{}%
\providecommand \url  [0]{\begingroup\@sanitize@url \@url }%
\providecommand \@url [1]{\endgroup\@href {#1}{\urlprefix }}%
\providecommand \urlprefix  [0]{URL }%
\providecommand \Eprint [0]{\href }%
\providecommand \doibase [0]{http://dx.doi.org/}%
\providecommand \selectlanguage [0]{\@gobble}%
\providecommand \bibinfo  [0]{\@secondoftwo}%
\providecommand \bibfield  [0]{\@secondoftwo}%
\providecommand \translation [1]{[#1]}%
\providecommand \BibitemOpen [0]{}%
\providecommand \bibitemStop [0]{}%
\providecommand \bibitemNoStop [0]{.\EOS\space}%
\providecommand \EOS [0]{\spacefactor3000\relax}%
\providecommand \BibitemShut  [1]{\csname bibitem#1\endcsname}%
\let\auto@bib@innerbib\@empty
\bibitem [{\citenamefont {Wilson}(1971{\natexlab{a}})}]{Wilson:1971bg}%
  \BibitemOpen
  \bibfield  {author} {\bibinfo {author} {\bibfnamefont {K.~G.}\ \bibnamefont
  {Wilson}},\ }\href {\doibase 10.1103/PhysRevB.4.3174} {\bibfield  {journal}
  {\bibinfo  {journal} {Phys. Rev.}\ }\textbf {\bibinfo {volume} {B4}},\
  \bibinfo {pages} {3174} (\bibinfo {year} {1971}{\natexlab{a}})}\BibitemShut
  {NoStop}%
\bibitem [{\citenamefont {Wilson}(1971{\natexlab{b}})}]{Wilson:1971dh}%
  \BibitemOpen
  \bibfield  {author} {\bibinfo {author} {\bibfnamefont {K.~G.}\ \bibnamefont
  {Wilson}},\ }\href {\doibase 10.1103/PhysRevB.4.3184} {\bibfield  {journal}
  {\bibinfo  {journal} {Phys. Rev.}\ }\textbf {\bibinfo {volume} {B4}},\
  \bibinfo {pages} {3184} (\bibinfo {year} {1971}{\natexlab{b}})}\BibitemShut
  {NoStop}%
\bibitem [{\citenamefont {Gross}\ and\ \citenamefont
  {Wilczek}(1973)}]{Gross:1973ju}%
  \BibitemOpen
  \bibfield  {author} {\bibinfo {author} {\bibfnamefont {D.~J.}\ \bibnamefont
  {Gross}}\ and\ \bibinfo {author} {\bibfnamefont {F.}~\bibnamefont
  {Wilczek}},\ }\href {\doibase 10.1103/PhysRevD.8.3633} {\bibfield  {journal}
  {\bibinfo  {journal} {Phys. Rev.}\ }\textbf {\bibinfo {volume} {D8}},\
  \bibinfo {pages} {3633} (\bibinfo {year} {1973})}\BibitemShut {NoStop}%
\bibitem [{\citenamefont {Politzer}(1973)}]{Politzer:1973fx}%
  \BibitemOpen
  \bibfield  {author} {\bibinfo {author} {\bibfnamefont {H.~D.}\ \bibnamefont
  {Politzer}},\ }\href {\doibase 10.1103/PhysRevLett.30.1346} {\bibfield
  {journal} {\bibinfo  {journal} {Phys. Rev. Lett.}\ }\textbf {\bibinfo
  {volume} {30}},\ \bibinfo {pages} {1346} (\bibinfo {year}
  {1973})}\BibitemShut {NoStop}%
\bibitem [{\citenamefont {Litim}\ and\ \citenamefont
  {Sannino}(2014)}]{Litim:2014uca}%
  \BibitemOpen
  \bibfield  {author} {\bibinfo {author} {\bibfnamefont {D.~F.}\ \bibnamefont
  {Litim}}\ and\ \bibinfo {author} {\bibfnamefont {F.}~\bibnamefont
  {Sannino}},\ }\href {\doibase 10.1007/JHEP12(2014)178} {\bibfield  {journal}
  {\bibinfo  {journal} {JHEP}\ }\textbf {\bibinfo {volume} {12}},\ \bibinfo
  {pages} {178} (\bibinfo {year} {2014})},\ \Eprint
  {http://arxiv.org/abs/1406.2337} {arXiv:1406.2337 [hep-th]} \BibitemShut
  {NoStop}%
\bibitem [{\citenamefont {Sannino}\ and\ \citenamefont
  {Skrinjar}(2018)}]{Sannino:2018suq}%
  \BibitemOpen
  \bibfield  {author} {\bibinfo {author} {\bibfnamefont {F.}~\bibnamefont
  {Sannino}}\ and\ \bibinfo {author} {\bibfnamefont {V.}~\bibnamefont
  {Skrinjar}},\ }\href@noop {} {\  (\bibinfo {year} {2018})},\ \Eprint
  {http://arxiv.org/abs/1802.10372} {arXiv:1802.10372 [hep-th]} \BibitemShut
  {NoStop}%
\bibitem [{\citenamefont {Litim}\ \emph {et~al.}(2016)\citenamefont {Litim},
  \citenamefont {Mojaza},\ and\ \citenamefont {Sannino}}]{Litim:2015iea}%
  \BibitemOpen
  \bibfield  {author} {\bibinfo {author} {\bibfnamefont {D.~F.}\ \bibnamefont
  {Litim}}, \bibinfo {author} {\bibfnamefont {M.}~\bibnamefont {Mojaza}}, \
  and\ \bibinfo {author} {\bibfnamefont {F.}~\bibnamefont {Sannino}},\ }\href
  {\doibase 10.1007/JHEP01(2016)081} {\bibfield  {journal} {\bibinfo  {journal}
  {JHEP}\ }\textbf {\bibinfo {volume} {01}},\ \bibinfo {pages} {081} (\bibinfo
  {year} {2016})},\ \Eprint {http://arxiv.org/abs/1501.03061} {arXiv:1501.03061
  [hep-th]} \BibitemShut {NoStop}%
\bibitem [{\citenamefont {Sannino}(2015)}]{Sannino:2015sel}%
  \BibitemOpen
  \bibfield  {author} {\bibinfo {author} {\bibfnamefont {F.}~\bibnamefont
  {Sannino}},\ }in\ \href
  {http://inspirehep.net/record/1407153/files/arXiv:1511.09022.pdf} {\emph
  {\bibinfo {booktitle} {{Proceedings, High-Precision alpha strong Measurements
  from LHC to FCC-ee: Geneva, Switzerland, October 2-13, 2015}}}}\ (\bibinfo
  {year} {2015})\ pp.\ \bibinfo {pages} {11--19},\ \Eprint
  {http://arxiv.org/abs/1511.09022} {arXiv:1511.09022 [hep-ph]} \BibitemShut
  {NoStop}%
\bibitem [{\citenamefont {Sannino}\ and\ \citenamefont
  {Shoemaker}(2015)}]{Sannino:2014lxa}%
  \BibitemOpen
  \bibfield  {author} {\bibinfo {author} {\bibfnamefont {F.}~\bibnamefont
  {Sannino}}\ and\ \bibinfo {author} {\bibfnamefont {I.~M.}\ \bibnamefont
  {Shoemaker}},\ }\href {\doibase 10.1103/PhysRevD.92.043518} {\bibfield
  {journal} {\bibinfo  {journal} {Phys. Rev.}\ }\textbf {\bibinfo {volume}
  {D92}},\ \bibinfo {pages} {043518} (\bibinfo {year} {2015})},\ \Eprint
  {http://arxiv.org/abs/1412.8034} {arXiv:1412.8034 [hep-ph]} \BibitemShut
  {NoStop}%
\bibitem [{\citenamefont {Abel}\ and\ \citenamefont
  {Sannino}(2017{\natexlab{a}})}]{Abel:2017ujy}%
  \BibitemOpen
  \bibfield  {author} {\bibinfo {author} {\bibfnamefont {S.}~\bibnamefont
  {Abel}}\ and\ \bibinfo {author} {\bibfnamefont {F.}~\bibnamefont {Sannino}},\
  }\href {\doibase 10.1103/PhysRevD.96.056028} {\bibfield  {journal} {\bibinfo
  {journal} {Phys. Rev.}\ }\textbf {\bibinfo {volume} {D96}},\ \bibinfo {pages}
  {056028} (\bibinfo {year} {2017}{\natexlab{a}})},\ \Eprint
  {http://arxiv.org/abs/1704.00700} {arXiv:1704.00700 [hep-ph]} \BibitemShut
  {NoStop}%
\bibitem [{\citenamefont {Abel}\ and\ \citenamefont
  {Sannino}(2017{\natexlab{b}})}]{Abel:2017rwl}%
  \BibitemOpen
  \bibfield  {author} {\bibinfo {author} {\bibfnamefont {S.}~\bibnamefont
  {Abel}}\ and\ \bibinfo {author} {\bibfnamefont {F.}~\bibnamefont {Sannino}},\
  }\href {\doibase 10.1103/PhysRevD.96.055021} {\bibfield  {journal} {\bibinfo
  {journal} {Phys. Rev.}\ }\textbf {\bibinfo {volume} {D96}},\ \bibinfo {pages}
  {055021} (\bibinfo {year} {2017}{\natexlab{b}})},\ \Eprint
  {http://arxiv.org/abs/1707.06638} {arXiv:1707.06638 [hep-ph]} \BibitemShut
  {NoStop}%
\bibitem [{\citenamefont {Pelaggi}\ \emph
  {et~al.}(2017{\natexlab{a}})\citenamefont {Pelaggi}, \citenamefont {Sannino},
  \citenamefont {Strumia},\ and\ \citenamefont {Vigiani}}]{Pelaggi:2017wzr}%
  \BibitemOpen
  \bibfield  {author} {\bibinfo {author} {\bibfnamefont {G.~M.}\ \bibnamefont
  {Pelaggi}}, \bibinfo {author} {\bibfnamefont {F.}~\bibnamefont {Sannino}},
  \bibinfo {author} {\bibfnamefont {A.}~\bibnamefont {Strumia}}, \ and\
  \bibinfo {author} {\bibfnamefont {E.}~\bibnamefont {Vigiani}},\ }\href
  {\doibase 10.3389/fphy.2017.00049} {\bibfield  {journal} {\bibinfo  {journal}
  {Front.in Phys.}\ }\textbf {\bibinfo {volume} {5}},\ \bibinfo {pages} {49}
  (\bibinfo {year} {2017}{\natexlab{a}})},\ \Eprint
  {http://arxiv.org/abs/1701.01453} {arXiv:1701.01453 [hep-ph]} \BibitemShut
  {NoStop}%
\bibitem [{\citenamefont {Mann}\ \emph {et~al.}(2017)\citenamefont {Mann},
  \citenamefont {Meffe}, \citenamefont {Sannino}, \citenamefont {Steele},
  \citenamefont {Wang},\ and\ \citenamefont {Zhang}}]{Mann:2017wzh}%
  \BibitemOpen
  \bibfield  {author} {\bibinfo {author} {\bibfnamefont {R.}~\bibnamefont
  {Mann}}, \bibinfo {author} {\bibfnamefont {J.}~\bibnamefont {Meffe}},
  \bibinfo {author} {\bibfnamefont {F.}~\bibnamefont {Sannino}}, \bibinfo
  {author} {\bibfnamefont {T.}~\bibnamefont {Steele}}, \bibinfo {author}
  {\bibfnamefont {Z.-W.}\ \bibnamefont {Wang}}, \ and\ \bibinfo {author}
  {\bibfnamefont {C.}~\bibnamefont {Zhang}},\ }\href {\doibase
  10.1103/PhysRevLett.119.261802} {\bibfield  {journal} {\bibinfo  {journal}
  {Phys. Rev. Lett.}\ }\textbf {\bibinfo {volume} {119}},\ \bibinfo {pages}
  {261802} (\bibinfo {year} {2017})},\ \Eprint
  {http://arxiv.org/abs/1707.02942} {arXiv:1707.02942 [hep-ph]} \BibitemShut
  {NoStop}%
\bibitem [{\citenamefont {Pelaggi}\ \emph
  {et~al.}(2017{\natexlab{b}})\citenamefont {Pelaggi}, \citenamefont
  {Plascencia}, \citenamefont {Salvio}, \citenamefont {Sannino}, \citenamefont
  {Smirnov},\ and\ \citenamefont {Strumia}}]{Pelaggi:2017abg}%
  \BibitemOpen
  \bibfield  {author} {\bibinfo {author} {\bibfnamefont {G.~M.}\ \bibnamefont
  {Pelaggi}}, \bibinfo {author} {\bibfnamefont {A.~D.}\ \bibnamefont
  {Plascencia}}, \bibinfo {author} {\bibfnamefont {A.}~\bibnamefont {Salvio}},
  \bibinfo {author} {\bibfnamefont {F.}~\bibnamefont {Sannino}}, \bibinfo
  {author} {\bibfnamefont {J.}~\bibnamefont {Smirnov}}, \ and\ \bibinfo
  {author} {\bibfnamefont {A.}~\bibnamefont {Strumia}},\ }\href@noop {} {\
  (\bibinfo {year} {2017}{\natexlab{b}})},\ \Eprint
  {http://arxiv.org/abs/1708.00437} {arXiv:1708.00437 [hep-ph]} \BibitemShut
  {NoStop}%
\bibitem [{\citenamefont {Bond}\ \emph {et~al.}(2017)\citenamefont {Bond},
  \citenamefont {Hiller}, \citenamefont {Kowalska},\ and\ \citenamefont
  {Litim}}]{Bond:2017wut}%
  \BibitemOpen
  \bibfield  {author} {\bibinfo {author} {\bibfnamefont {A.~D.}\ \bibnamefont
  {Bond}}, \bibinfo {author} {\bibfnamefont {G.}~\bibnamefont {Hiller}},
  \bibinfo {author} {\bibfnamefont {K.}~\bibnamefont {Kowalska}}, \ and\
  \bibinfo {author} {\bibfnamefont {D.~F.}\ \bibnamefont {Litim}},\ }\href
  {\doibase 10.1007/JHEP08(2017)004} {\bibfield  {journal} {\bibinfo  {journal}
  {JHEP}\ }\textbf {\bibinfo {volume} {08}},\ \bibinfo {pages} {004} (\bibinfo
  {year} {2017})},\ \Eprint {http://arxiv.org/abs/1702.01727} {arXiv:1702.01727
  [hep-ph]} \BibitemShut {NoStop}%
\bibitem [{\citenamefont {Eichhorn}\ \emph
  {et~al.}(2018{\natexlab{a}})\citenamefont {Eichhorn}, \citenamefont {Held},\
  and\ \citenamefont {Griend}}]{Eichhorn:2018vah}%
  \BibitemOpen
  \bibfield  {author} {\bibinfo {author} {\bibfnamefont {A.}~\bibnamefont
  {Eichhorn}}, \bibinfo {author} {\bibfnamefont {A.}~\bibnamefont {Held}}, \
  and\ \bibinfo {author} {\bibfnamefont {P.~V.}\ \bibnamefont {Griend}},\
  }\href@noop {} {\  (\bibinfo {year} {2018}{\natexlab{a}})},\ \Eprint
  {http://arxiv.org/abs/1802.08589} {arXiv:1802.08589 [hep-ph]} \BibitemShut
  {NoStop}%
\bibitem [{\citenamefont {Eichhorn}\ \emph {et~al.}(2017)\citenamefont
  {Eichhorn}, \citenamefont {Held},\ and\ \citenamefont
  {Wetterich}}]{Eichhorn:2017muy}%
  \BibitemOpen
  \bibfield  {author} {\bibinfo {author} {\bibfnamefont {A.}~\bibnamefont
  {Eichhorn}}, \bibinfo {author} {\bibfnamefont {A.}~\bibnamefont {Held}}, \
  and\ \bibinfo {author} {\bibfnamefont {C.}~\bibnamefont {Wetterich}},\
  }\href@noop {} {\  (\bibinfo {year} {2017})},\ \Eprint
  {http://arxiv.org/abs/1711.02949} {arXiv:1711.02949 [hep-th]} \BibitemShut
  {NoStop}%
\bibitem [{\citenamefont {Reichert}\ \emph {et~al.}(2017)\citenamefont
  {Reichert}, \citenamefont {Eichhorn}, \citenamefont {Gies}, \citenamefont
  {Pawlowski}, \citenamefont {Plehn},\ and\ \citenamefont
  {Scherer}}]{Reichert:2017puo}%
  \BibitemOpen
  \bibfield  {author} {\bibinfo {author} {\bibfnamefont {M.}~\bibnamefont
  {Reichert}}, \bibinfo {author} {\bibfnamefont {A.}~\bibnamefont {Eichhorn}},
  \bibinfo {author} {\bibfnamefont {H.}~\bibnamefont {Gies}}, \bibinfo {author}
  {\bibfnamefont {J.~M.}\ \bibnamefont {Pawlowski}}, \bibinfo {author}
  {\bibfnamefont {T.}~\bibnamefont {Plehn}}, \ and\ \bibinfo {author}
  {\bibfnamefont {M.~M.}\ \bibnamefont {Scherer}},\ }\href@noop {} {\
  (\bibinfo {year} {2017})},\ \Eprint {http://arxiv.org/abs/1711.00019}
  {arXiv:1711.00019 [hep-ph]} \BibitemShut {NoStop}%
\bibitem [{\citenamefont {Eichhorn}\ \emph
  {et~al.}(2018{\natexlab{b}})\citenamefont {Eichhorn}, \citenamefont
  {Lippoldt},\ and\ \citenamefont {Skrinjar}}]{Eichhorn:2017sok}%
  \BibitemOpen
  \bibfield  {author} {\bibinfo {author} {\bibfnamefont {A.}~\bibnamefont
  {Eichhorn}}, \bibinfo {author} {\bibfnamefont {S.}~\bibnamefont {Lippoldt}},
  \ and\ \bibinfo {author} {\bibfnamefont {V.}~\bibnamefont {Skrinjar}},\
  }\href {\doibase 10.1103/PhysRevD.97.026002} {\bibfield  {journal} {\bibinfo
  {journal} {Phys. Rev.}\ }\textbf {\bibinfo {volume} {D97}},\ \bibinfo {pages}
  {026002} (\bibinfo {year} {2018}{\natexlab{b}})},\ \Eprint
  {http://arxiv.org/abs/1710.03005} {arXiv:1710.03005 [hep-th]} \BibitemShut
  {NoStop}%
\bibitem [{\citenamefont {Eichhorn}\ and\ \citenamefont
  {Versteegen}(2018)}]{Eichhorn:2017lry}%
  \BibitemOpen
  \bibfield  {author} {\bibinfo {author} {\bibfnamefont {A.}~\bibnamefont
  {Eichhorn}}\ and\ \bibinfo {author} {\bibfnamefont {F.}~\bibnamefont
  {Versteegen}},\ }\href {\doibase 10.1007/JHEP01(2018)030} {\bibfield
  {journal} {\bibinfo  {journal} {JHEP}\ }\textbf {\bibinfo {volume} {01}},\
  \bibinfo {pages} {030} (\bibinfo {year} {2018})},\ \Eprint
  {http://arxiv.org/abs/1709.07252} {arXiv:1709.07252 [hep-th]} \BibitemShut
  {NoStop}%
\bibitem [{\citenamefont {Intriligator}\ and\ \citenamefont
  {Wecht}(2003)}]{Intriligator:2003jj}%
  \BibitemOpen
  \bibfield  {author} {\bibinfo {author} {\bibfnamefont {K.~A.}\ \bibnamefont
  {Intriligator}}\ and\ \bibinfo {author} {\bibfnamefont {B.}~\bibnamefont
  {Wecht}},\ }\href {\doibase 10.1016/S0550-3213(03)00459-0} {\bibfield
  {journal} {\bibinfo  {journal} {Nucl. Phys.}\ }\textbf {\bibinfo {volume}
  {B667}},\ \bibinfo {pages} {183} (\bibinfo {year} {2003})},\ \Eprint
  {http://arxiv.org/abs/hep-th/0304128} {arXiv:hep-th/0304128 [hep-th]}
  \BibitemShut {NoStop}%
\bibitem [{\citenamefont {Hofman}\ and\ \citenamefont
  {Maldacena}(2008)}]{Hofman:2008ar}%
  \BibitemOpen
  \bibfield  {author} {\bibinfo {author} {\bibfnamefont {D.~M.}\ \bibnamefont
  {Hofman}}\ and\ \bibinfo {author} {\bibfnamefont {J.}~\bibnamefont
  {Maldacena}},\ }\href {\doibase 10.1088/1126-6708/2008/05/012} {\bibfield
  {journal} {\bibinfo  {journal} {JHEP}\ }\textbf {\bibinfo {volume} {05}},\
  \bibinfo {pages} {012} (\bibinfo {year} {2008})},\ \Eprint
  {http://arxiv.org/abs/0803.1467} {arXiv:0803.1467 [hep-th]} \BibitemShut
  {NoStop}%
\bibitem [{\citenamefont {Intriligator}\ and\ \citenamefont
  {Sannino}(2015)}]{Intriligator:2015xxa}%
  \BibitemOpen
  \bibfield  {author} {\bibinfo {author} {\bibfnamefont {K.}~\bibnamefont
  {Intriligator}}\ and\ \bibinfo {author} {\bibfnamefont {F.}~\bibnamefont
  {Sannino}},\ }\href {\doibase 10.1007/JHEP11(2015)023} {\bibfield  {journal}
  {\bibinfo  {journal} {JHEP}\ }\textbf {\bibinfo {volume} {11}},\ \bibinfo
  {pages} {023} (\bibinfo {year} {2015})},\ \Eprint
  {http://arxiv.org/abs/1508.07411} {arXiv:1508.07411 [hep-th]} \BibitemShut
  {NoStop}%
\bibitem [{\citenamefont {Martin}\ and\ \citenamefont
  {Wells}(2001)}]{Martin:2000cr}%
  \BibitemOpen
  \bibfield  {author} {\bibinfo {author} {\bibfnamefont {S.~P.}\ \bibnamefont
  {Martin}}\ and\ \bibinfo {author} {\bibfnamefont {J.~D.}\ \bibnamefont
  {Wells}},\ }\href {\doibase 10.1103/PhysRevD.64.036010} {\bibfield  {journal}
  {\bibinfo  {journal} {Phys. Rev.}\ }\textbf {\bibinfo {volume} {D64}},\
  \bibinfo {pages} {036010} (\bibinfo {year} {2001})},\ \Eprint
  {http://arxiv.org/abs/hep-ph/0011382} {arXiv:hep-ph/0011382 [hep-ph]}
  \BibitemShut {NoStop}%
\bibitem [{\citenamefont {Bajc}\ and\ \citenamefont
  {Sannino}(2016)}]{Bajc:2016efj}%
  \BibitemOpen
  \bibfield  {author} {\bibinfo {author} {\bibfnamefont {B.}~\bibnamefont
  {Bajc}}\ and\ \bibinfo {author} {\bibfnamefont {F.}~\bibnamefont {Sannino}},\
  }\href {\doibase 10.1007/JHEP12(2016)141} {\bibfield  {journal} {\bibinfo
  {journal} {JHEP}\ }\textbf {\bibinfo {volume} {12}},\ \bibinfo {pages} {141}
  (\bibinfo {year} {2016})},\ \Eprint {http://arxiv.org/abs/1610.09681}
  {arXiv:1610.09681 [hep-th]} \BibitemShut {NoStop}%
\bibitem [{\citenamefont {Bajc}\ \emph {et~al.}(2017)\citenamefont {Bajc},
  \citenamefont {Dondi},\ and\ \citenamefont {Sannino}}]{Bajc:2017xwx}%
  \BibitemOpen
  \bibfield  {author} {\bibinfo {author} {\bibfnamefont {B.}~\bibnamefont
  {Bajc}}, \bibinfo {author} {\bibfnamefont {N.~A.}\ \bibnamefont {Dondi}}, \
  and\ \bibinfo {author} {\bibfnamefont {F.}~\bibnamefont {Sannino}},\
  }\href@noop {} {\  (\bibinfo {year} {2017})},\ \Eprint
  {http://arxiv.org/abs/1709.07436} {arXiv:1709.07436 [hep-th]} \BibitemShut
  {NoStop}%
\bibitem [{\citenamefont {Palanques-Mestre}\ and\ \citenamefont
  {Pascual}(1984)}]{PalanquesMestre:1983zy}%
  \BibitemOpen
  \bibfield  {author} {\bibinfo {author} {\bibfnamefont {A.}~\bibnamefont
  {Palanques-Mestre}}\ and\ \bibinfo {author} {\bibfnamefont {P.}~\bibnamefont
  {Pascual}},\ }\href {\doibase 10.1007/BF01212398} {\bibfield  {journal}
  {\bibinfo  {journal} {Commun. Math. Phys.}\ }\textbf {\bibinfo {volume}
  {95}},\ \bibinfo {pages} {277} (\bibinfo {year} {1984})}\BibitemShut
  {NoStop}%
\bibitem [{\citenamefont {Gracey}(1996)}]{Gracey:1996he}%
  \BibitemOpen
  \bibfield  {author} {\bibinfo {author} {\bibfnamefont {J.~A.}\ \bibnamefont
  {Gracey}},\ }\href {\doibase 10.1016/0370-2693(96)00105-0} {\bibfield
  {journal} {\bibinfo  {journal} {Phys. Lett.}\ }\textbf {\bibinfo {volume}
  {B373}},\ \bibinfo {pages} {178} (\bibinfo {year} {1996})},\ \Eprint
  {http://arxiv.org/abs/hep-ph/9602214} {arXiv:hep-ph/9602214 [hep-ph]}
  \BibitemShut {NoStop}%
\bibitem [{\citenamefont {Holdom}(2011)}]{Holdom:2010qs}%
  \BibitemOpen
  \bibfield  {author} {\bibinfo {author} {\bibfnamefont {B.}~\bibnamefont
  {Holdom}},\ }\href {\doibase 10.1016/j.physletb.2010.09.037} {\bibfield
  {journal} {\bibinfo  {journal} {Phys. Lett.}\ }\textbf {\bibinfo {volume}
  {B694}},\ \bibinfo {pages} {74} (\bibinfo {year} {2011})},\ \Eprint
  {http://arxiv.org/abs/1006.2119} {arXiv:1006.2119 [hep-ph]} \BibitemShut
  {NoStop}%
\bibitem [{\citenamefont {Pica}\ and\ \citenamefont
  {Sannino}(2011)}]{Pica:2010xq}%
  \BibitemOpen
  \bibfield  {author} {\bibinfo {author} {\bibfnamefont {C.}~\bibnamefont
  {Pica}}\ and\ \bibinfo {author} {\bibfnamefont {F.}~\bibnamefont {Sannino}},\
  }\href {\doibase 10.1103/PhysRevD.83.035013} {\bibfield  {journal} {\bibinfo
  {journal} {Phys. Rev.}\ }\textbf {\bibinfo {volume} {D83}},\ \bibinfo {pages}
  {035013} (\bibinfo {year} {2011})},\ \Eprint {http://arxiv.org/abs/1011.5917}
  {arXiv:1011.5917 [hep-ph]} \BibitemShut {NoStop}%
\bibitem [{\citenamefont {Shrock}(2014)}]{Shrock:2013cca}%
  \BibitemOpen
  \bibfield  {author} {\bibinfo {author} {\bibfnamefont {R.}~\bibnamefont
  {Shrock}},\ }\href {\doibase 10.1103/PhysRevD.89.045019} {\bibfield
  {journal} {\bibinfo  {journal} {Phys. Rev.}\ }\textbf {\bibinfo {volume}
  {D89}},\ \bibinfo {pages} {045019} (\bibinfo {year} {2014})},\ \Eprint
  {http://arxiv.org/abs/1311.5268} {arXiv:1311.5268 [hep-th]} \BibitemShut
  {NoStop}%
\bibitem [{\citenamefont {Antipin}\ and\ \citenamefont
  {Sannino}(2017)}]{Antipin:2017ebo}%
  \BibitemOpen
  \bibfield  {author} {\bibinfo {author} {\bibfnamefont {O.}~\bibnamefont
  {Antipin}}\ and\ \bibinfo {author} {\bibfnamefont {F.}~\bibnamefont
  {Sannino}},\ }\href@noop {} {\  (\bibinfo {year} {2017})},\ \Eprint
  {http://arxiv.org/abs/1709.02354} {arXiv:1709.02354 [hep-ph]} \BibitemShut
  {NoStop}%
\bibitem [{\citenamefont {Sannino}(2009)}]{Sannino:2009za}%
  \BibitemOpen
  \bibfield  {author} {\bibinfo {author} {\bibfnamefont {F.}~\bibnamefont
  {Sannino}},\ }\bibfield  {booktitle} {\emph {\bibinfo {booktitle}
  {{Non-perturbative gravity and quantum chromodynamics. Proceedings, 49th
  Cracow School of Theoretical Physics, Zakopane, Poland, May 31-June 10,
  2009}}},\ }\href@noop {} {\bibfield  {journal} {\bibinfo  {journal} {Acta
  Phys. Polon.}\ }\textbf {\bibinfo {volume} {B40}},\ \bibinfo {pages} {3533}
  (\bibinfo {year} {2009})},\ \Eprint {http://arxiv.org/abs/0911.0931}
  {arXiv:0911.0931 [hep-ph]} \BibitemShut {NoStop}%
\bibitem [{\citenamefont {Pica}(2016)}]{Pica:2017gcb}%
  \BibitemOpen
  \bibfield  {author} {\bibinfo {author} {\bibfnamefont {C.}~\bibnamefont
  {Pica}},\ }\bibfield  {booktitle} {\emph {\bibinfo {booktitle} {{Proceedings,
  34th International Symposium on Lattice Field Theory (Lattice 2016):
  Southampton, UK, July 24-30, 2016}}},\ }\href@noop {} {\bibfield  {journal}
  {\bibinfo  {journal} {PoS}\ }\textbf {\bibinfo {volume} {LATTICE2016}},\
  \bibinfo {pages} {015} (\bibinfo {year} {2016})},\ \Eprint
  {http://arxiv.org/abs/1701.07782} {arXiv:1701.07782 [hep-lat]} \BibitemShut
  {NoStop}%
\bibitem [{\citenamefont {Kowalska}\ and\ \citenamefont
  {Sessolo}(2017)}]{Kowalska:2017pkt}%
  \BibitemOpen
  \bibfield  {author} {\bibinfo {author} {\bibfnamefont {K.}~\bibnamefont
  {Kowalska}}\ and\ \bibinfo {author} {\bibfnamefont {E.~M.}\ \bibnamefont
  {Sessolo}},\ }\href@noop {} {\  (\bibinfo {year} {2017})},\ \Eprint
  {http://arxiv.org/abs/1712.06859} {arXiv:1712.06859 [hep-ph]} \BibitemShut
  {NoStop}%
\bibitem [{\citenamefont {Ferreira}\ \emph
  {et~al.}(1997{\natexlab{a}})\citenamefont {Ferreira}, \citenamefont {Jack},\
  and\ \citenamefont {Jones}}]{Ferreira:1997rc}%
  \BibitemOpen
  \bibfield  {author} {\bibinfo {author} {\bibfnamefont {P.~M.}\ \bibnamefont
  {Ferreira}}, \bibinfo {author} {\bibfnamefont {I.}~\bibnamefont {Jack}}, \
  and\ \bibinfo {author} {\bibfnamefont {D.~R.~T.}\ \bibnamefont {Jones}},\
  }\href {\doibase 10.1016/S0370-2693(97)00291-8} {\bibfield  {journal}
  {\bibinfo  {journal} {Phys. Lett.}\ }\textbf {\bibinfo {volume} {B399}},\
  \bibinfo {pages} {258} (\bibinfo {year} {1997}{\natexlab{a}})},\ \Eprint
  {http://arxiv.org/abs/hep-ph/9702304} {arXiv:hep-ph/9702304 [hep-ph]}
  \BibitemShut {NoStop}%
\bibitem [{\citenamefont {Ferreira}\ \emph
  {et~al.}(1997{\natexlab{b}})\citenamefont {Ferreira}, \citenamefont {Jack},
  \citenamefont {Jones},\ and\ \citenamefont {North}}]{Ferreira:1997bi}%
  \BibitemOpen
  \bibfield  {author} {\bibinfo {author} {\bibfnamefont {P.~M.}\ \bibnamefont
  {Ferreira}}, \bibinfo {author} {\bibfnamefont {I.}~\bibnamefont {Jack}},
  \bibinfo {author} {\bibfnamefont {D.~R.~T.}\ \bibnamefont {Jones}}, \ and\
  \bibinfo {author} {\bibfnamefont {C.~G.}\ \bibnamefont {North}},\ }\href
  {\doibase 10.1016/S0550-3213(97)00448-3} {\bibfield  {journal} {\bibinfo
  {journal} {Nucl. Phys.}\ }\textbf {\bibinfo {volume} {B504}},\ \bibinfo
  {pages} {108} (\bibinfo {year} {1997}{\natexlab{b}})},\ \Eprint
  {http://arxiv.org/abs/hep-ph/9705328} {arXiv:hep-ph/9705328 [hep-ph]}
  \BibitemShut {NoStop}%
\bibitem [{\citenamefont {Mihaila}(2013)}]{Mihaila:2014caa}%
  \BibitemOpen
  \bibfield  {author} {\bibinfo {author} {\bibfnamefont {L.}~\bibnamefont
  {Mihaila}},\ }\bibfield  {booktitle} {\emph {\bibinfo {booktitle}
  {{Proceedings, 11th International Symposium on Radiative Corrections
  "Application of Quantum Field Theory to Phenomenology" (RADCOR 2013): Durham,
  UK, September 22-27, 2013}}},\ }\href@noop {} {\bibfield  {journal} {\bibinfo
   {journal} {PoS}\ }\textbf {\bibinfo {volume} {RADCOR2013}},\ \bibinfo
  {pages} {060} (\bibinfo {year} {2013})}\BibitemShut {NoStop}%
\bibitem [{\citenamefont {Esbensen}\ \emph {et~al.}(2016)\citenamefont
  {Esbensen}, \citenamefont {Ryttov},\ and\ \citenamefont
  {Sannino}}]{Esbensen:2015cjw}%
  \BibitemOpen
  \bibfield  {author} {\bibinfo {author} {\bibfnamefont {J.~K.}\ \bibnamefont
  {Esbensen}}, \bibinfo {author} {\bibfnamefont {T.~A.}\ \bibnamefont
  {Ryttov}}, \ and\ \bibinfo {author} {\bibfnamefont {F.}~\bibnamefont
  {Sannino}},\ }\href {\doibase 10.1103/PhysRevD.93.045009} {\bibfield
  {journal} {\bibinfo  {journal} {Phys. Rev.}\ }\textbf {\bibinfo {volume}
  {D93}},\ \bibinfo {pages} {045009} (\bibinfo {year} {2016})},\ \Eprint
  {http://arxiv.org/abs/1512.04402} {arXiv:1512.04402 [hep-th]} \BibitemShut
  {NoStop}%
\end{thebibliography}%

\end{document}